\documentclass[reqno,11pt]{amsart}
\usepackage[utf8]{inputenc}
\usepackage{enumitem}
\usepackage[linktocpage]{hyperref}
\usepackage{graphicx}
\usepackage{amscd}
\usepackage{slashed}
\usepackage{amssymb}
\usepackage{mathtools} 
\usepackage{pstricks}
\usepackage[mathscr]{eucal}
\numberwithin{equation}{section}
\allowdisplaybreaks[1]
\definecolor{labelkey}{gray}{.65}

\title[Causal Fermion Systems and Octonions]{Causal Fermion Systems and Octonions}

\author[F.\ Finster]{Felix Finster}
\address{Fakult\"at f\"ur Mathematik \\ Universit\"at Regensburg \\ D-93040 Regensburg \\ Germany}%
\email{finster@ur.de}

\author[N.G.\ Gresnigt]{Niels G. Gresnigt}
\address{Department of Physics \\ Xi'an Jiaotong-Liverpool University \\
111 Ren’ai Road \\ Suzhou Industrial Park \\ Suzhou \\ Jiangsu Province \\ P. R. China 215123}
\email{niels.gresnigt@xjtlu.edu.cn}

\author[J.M.\ Isidro]{Jos{\'e} M. Isidro}
\address{Instituto Universitario de Matem\'atica Pura y Aplicada \\ Universidad Polit\'ecnica de Valencia \\
Valencia 46022 \\ Spain}
\email{joissan@mat.upv.es}

\author[A.\ Marcian{\`o}]{Antonino Marcian{\`o}}
\address{Department of Physics, Fudan University \\ Department of Physics (office n. S234) \\
n. 2005 Songhu Road \\ Shanghai 200438 \\ China}
\address{Laboratori Nazionali di Frascati INFN \\ 54 Via Enrico Fermi \\ Frascati 00044 \\ Italy}
\email{marciano@fudan.edu.cn}

\author[C.F.\ Paganini]{Claudio F. Paganini}
\address{Fakult\"at f\"ur Mathematik \\ Universit\"at Regensburg \\ D-93040 Regensburg \\ Germany}%
\email{claudio.paganini@ur.de}

\author[T.P.\ Singh]{Tejinder P. Singh \\ \\ March 2024}
\address{Inter-University Centre for Astronomy and Astrophysics,
Post Bag 4, Ganeshkhind, Pune 411007, India and
Tata Institute of Fundamental Research, Homi Bhabha Road, Mumbai 400005, India}
\email{tejinder.singh@iucaa.in; tpsingh@tifr.res.in}

\newtheorem{Def}{Definition}[section]

\newcommand{\Thanks}{\vspace*{.5em} \noindent \thanks}
\newcommand{\beq}{\begin{equation}}
\newcommand{\eeq}{\end{equation}}
\newcommand{\Proof}{\begin{proof}}
\newcommand{\QED}{\end{proof} \noindent}

\newcommand{\la}{\langle}
\newcommand{\ra}{\rangle}
\newcommand{\bra}{\mathopen{<}}
\newcommand{\ket}{\mathclose{>}}
\newcommand{\Sl}{\mathopen{\prec}}
\newcommand{\Sr}{\mathclose{\succ}}

\newcommand{\C}{\mathbb{C}}
\newcommand{\R}{\mathbb{R}}
\newcommand{\1}{\mbox{\rm 1 \hspace{-1.05 em} 1}}

\newcommand{\N}{\mathbb{N}}
\newcommand{\Pdd}{\mbox{$\partial$ \hspace{-1.2 em} $/$}}

\renewcommand{\H}{\mathscr{H}}
\newcommand{\U}{{\rm{U}}}
\newcommand{\SU}{{\rm{SU}}}

\newcommand{\bep}{\begin{pmatrix}}
\newcommand{\enp}{\end{pmatrix}}

\renewcommand{\O}{\mathscr{O}}

\newcommand{\F}{{\mathscr{F}}}

\newcommand{\B}{{\mathscr{B}}}
\renewcommand{\O}{{\mathscr{O}}}
\renewcommand{\L}{{\mathcal{L}}}
\newcommand{\Sact}{{\mathcal{S}}}
\newcommand{\s}{{\mathfrak{s}}}

\newcommand{\Lin}{\text{\rm{L}}}

\newcommand{\sea}{\text{\rm{sea}}}

\DeclareMathOperator{\Spin}{Spin}

\newcommand{\reg}{\text{\rm{reg}}}
\newcommand{\lec}{\text{\rm{le}}}
\newcommand{\hec}{\text{\rm{he}}}

\newcommand{\A}{\myscr A}
\renewcommand{\imath}{\rm{i}}

\DeclareFontFamily{OT1}{rsfso}{}
\DeclareFontShape{OT1}{rsfso}{m}{n}{ <-7> rsfso5 <7-10> rsfso7 <10-> rsfso10}{}
\DeclareMathAlphabet{\myscr}{OT1}{rsfso}{m}{n}

\newcommand\Felix[1]{}

\DeclareMathOperator{\Tr}{Tr}
\DeclareMathOperator{\tr}{tr}

\DeclareMathOperator{\supp}{supp}

\newcommand{\bitem}{\begin{itemize}[leftmargin=2em]}
\newcommand{\eitem}{\end{itemize}}
\newcommand{\itemD}{\item[{\raisebox{0.125em}{\tiny $\blacktriangleright$}}]}

\newcommand{\bb}[1]{\mathbb{#1}}

\begin{document}

\maketitle

\begin{abstract}
We compare the structures and methods in the theory of causal fermion systems
with approaches to fundamental physics based on division algebras, in particular the octonions.
We find that octonions and, more generally, tensor products of division algebras come up naturally to describe
the symmetries of the vacuum configuration of a causal fermion system.
This is achieved by associating the real and imaginary octonion basis elements with the neutrino and charged sectors of the vacuum fermionic projector, respectively. Conversely,
causal fermion systems provide octonionic theories with
spacetime structures and dynamical equations via the causal action principle.
In this way, octonionic theories and causal fermion systems complement each other.
\end{abstract}

\tableofcontents

\section{Introduction} \label{secintro}
This article is part of a series of papers comparing the structures and ideas of different approaches to fundamental physics that was started with~\cite{eth-cfs, mmt-cfs} and will be continued in~\cite{cfstdqg, tracedynamics}. Each of these papers contains an introduction to the theories under consideration including an overview of their accomplishments. These overviews serve as a starting point for a reader familiar with one of the approaches to engage with the other approach. The papers proceed with a detailed comparison of these two theories. Ultimately, the goal is to motivate the community to establish an extensive collection of such articles as a sort of ``Rosetta stone'' for approaches to fundamental physics. The hope is that such a set of dictionaries facilitates the exchange of ideas across approaches and thereby catalyzes progress in the foundations of physics. This addresses a distinctly different goal from overview articles such as~\cite{loll2022quantum, Mielczarek_2018, deBoer:2824341} or~\cite{addazi2022quantum} that try to cover the development across many approaches simultaneously. By focusing on two sets of ideas at a time, a greater level of depth can be achieved and more specific issues can be addressed.

The present article resulted from discussions of the authors with the wish of connecting
two different approaches to fundamental physics:
The approach relating {\em{octonions}} (or division algebras in general, although we restrict ourselves in the present articles to octonions and use the term octonions and division algebras synonymously) with fundamental physics on one side and
{\em{causal fermion systems}} on the other. Our report uncovers many surprising similarities and analogies, but also
points to differences and discusses open research questions.
One of our key findings is that octonions (more generally, tensor products of division algebras) come up naturally to describe symmetries or approximate
symmetries of the vacuum configuration of a causal fermion system.
Likewise, the interactions of a causal fermion system can be described in terms of
algebras of endomorphisms acting on the octonions.
In this way, the approaches connecting division algebras, in particular octonions, to the standard model
(see for example~\cite{dixon2013division,furey2016standard,todorov-dubois-violette,manogue-dray-wilson}) may give a more fundamental understanding for the structure of the vacuum of the
causal fermion system and for the form of possible interactions.
On the other hand, the theory of causal fermion systems may provide the octonionic theories
with additional structures needed for developing them into full physical theories.
In particular, causal fermion systems give rise to spacetime structures (like causality and geometry)
and equations describing the dynamics (via the causal action principle).
In this way, the octonionic theories and causal fermion systems complement each other.
Moreover, there are interesting and inspiring connections between the ideas
underlying the approaches. Studying these connections further should, on the one hand, provide
a dynamical framework for division algebraic descriptions of the standard model
and, on the other hand, give restrictions for possible matter models which can be realized as causal fermion systems.

Since the pioneering work of G\"uynadin and G\"ursey in the 1970s describing the~$\SU(3)$ color symmetry of quarks using octonions, attempts using octonions, together with the remaining three normed division algebras, to explain the structure, gauge groups, and particle multiplets of the standard model have grown in popularity. The extensive works of Dixon, culminated in the book~\cite{dixon2013division}, demonstrate how many of the mathematical features of the standard model, including its gauge symmetries to which a single generation of fermions and leptons are subject, together with the correct multiplets into which they fall, are inherent in the 64 real dimensional composition algebra
$\bb{T}=\mathbb{R}\otimes\mathbb{C}\otimes\mathbb{H}\otimes\mathbb{O}$. In a closely related approach, Furey focuses on the Clifford algebras which are generated by
linear operators acting on (tensor products of) division algebras via composition.
She describes the particle multiplets in terms of minimal left ideals of these algebras and the gauge groups as the subgroups of the Clifford algebra's spin group that preserve the structure of these minimal left ideals~\cite{furey2016standard,furey231,furey232}. Many other authors, not cited here, have contributed to these and related approaches. Such division algebraic constructions offer a number of attractive features, most notably a derivation of the standard model gauge groups from a minimal mathematical framework, together with the correct chiral fermion representations.

The theory of {\em{causal fermion systems}} is a recent approach to fundamental physics
(see the basics in Section~\ref{seccfsintro}, the reviews~\cite{dice2014, nrstg, review}, the textbooks~\cite{cfs, intro}
or the website~\cite{cfsweblink}).
In this approach, spacetime and all objects therein are described by a measure~$\rho$
on a set~$\F$ of linear operators on a Hilbert space~$(\H, \la .|. \ra_\H)$. 
The physical equations are formulated by means of the so-called {\em{causal action principle}},
a nonlinear variational principle where an action~$\Sact$ is minimized under variations of the measure~$\rho$.
In different limiting cases, causal fermion systems give rise to the standard model of particle physics
and gravity on the level of classical field theory~\cite{cfs} as well as to quantum field
theory~\cite{fockfermionic, fockentangle, fockdynamics}.

The connection between causal fermion systems and octonionic theories is established
by analyzing the symmetry properties of the causal fermion system describing the Minkowski vacuum.
In order to describe the Minkowski vacuum by a causal fermion system, one needs to specify the
configuration of the fermions. By doing this ad hoc in a specific way, 
the causal action principle yields the gauge groups and couplings of the standard model
(for more details see Section~\ref{seccfssm}).
As we shall see, the vacuum configuration can be described naturally using octonions.
Likewise, the symmetries are described by the associative multiplication algebras generated from the octonions.
Even more, the interaction as described by the causal action principle can be written
using automorphism groups of octonions and the exceptional Jordan algebra.
In this manner, one uncovers close connections between causal fermion systems and octonionic theories.
The main part of this paper is devoted to working out these connections in detail
(see Section~\ref{sec:comparison} and Section~\ref{sece68}).
A related question is whether octonions provide a better and more systematic understanding for
how the vacuum is to be chosen. Here we explore the possibility that the chiral asymmetry is no
longer built in from the beginning, but instead arises dynamically by a chiral symmetry breaking effect
(Section~\ref{secchiralsymm}). Analyzing these questions in more detail should also give new insight into
the rigidity of the particle model in the theory of causal fermion systems.

The paper is organized as follows. We begin by giving short but self-contained reviews
of octonionic theories (Section~\ref{secintroocto}) and to causal fermion systems (Section~\ref{seccfsintro}).
In Section~\ref{sec:comparison} we compare these approaches.
Section~\ref{secoutlook} gives an outlook and spells out several directions for future research.
Moreover, as Majorana spinors are important for certain aspects of octonionic theories,
in Appendix~\ref{appmajorana} we explain how Majorana spinors can be described by a causal fermion system.

\section{Division Algebras and the Standard Model} \label{secintroocto}
\subsection{Motivation}
The standard model of particle physics is a gauge theory based on the group~$\SU(3)\times \SU(2)\times \U(1)$ with particle multiplets falling into irreducible representations of these Lie groups. Despite its undeniable success, the standard model has some striking theoretical and conceptual puzzles that remain unanswered. For example:
\bitem
\itemD  What is the origin of the standard model gauge group, and why do the gauge groups, as well as the spacetime symmetries appear as separate factors?
\itemD Why do elementary fermions fall into the observed multiplets (irreducible representations of the Lie groups), and not others?
\itemD Why is~$\SU(2)$ chiral and broken whereas~$\SU(3)$ is non-chiral and exact?
\itemD Why are there three generations of fermions?
\itemD The theoretical origins of electroweak symmetry breaking and of CP violation are unknown.
\eitem
These, among other, unresolved puzzles provide a hint that the standard model likely constitutes an approximation to a yet more fundamental theory.

Grand unified theories (GUTs) attempt to address some of these issues by unifying the separate gauge groups into one larger group. A classic example involves the packaging of~$\SU(3)$, $\SU(2)$ and~$\U(1)$ into~$\SU(5)$. However, with an infinite number of  Lie groups at one's disposal, how does one pick those relevant to particle physics? Furthermore, having selected the groups of interest, how to choose from the infinitely many multiplets and representations?

A more discerning approach is to instead search for a simple fundamental mathematical structure from which the symmetries of the standard model and the right representations arise naturally. Nature only admits four normed division algebras (over the field of real numbers): the real numbers~$\mathbb{R}$ (1D), complex numbers~$\mathbb{C}$ (2D), the non-commutative quaternions~$\mathbb{H}$ (4D), and finally the non-associative octonions~$\mathbb{O}$ (8D). These four algebras are extremely generative, and the existence of the four division algebras gives rise to all (classical and exceptional) Lie groups. Additionally, all Clifford and Jordan algebras correspond to matrix algebras over the division algebras.

The (hyper)complex division algebras also happen to be intimately related to the three Lie gauge groups that appear in the standard model: The imaginary element of~$\bb{C}$ generates~$\U(1)$ via exponentiation; those of~$\bb{H}$
generate~$\SU(2)$; whereas the automorphism group of~$\bb{O}$, $G_2$, has an~$\SU(3)$ subgroup that is the stability group of a fixed imaginary direction.

Not surprisingly therefore, a growing number of authors over the years have sought to  connect the normed division algebras, in particular the octonions, to the gauge symmetries and particle multiplets of the standard model.

\subsection{Historical Overview}\label{historical}

The octonions first appeared in relation to quantum theory in the classification of Jordan algebras by Jordan et.\ al.~\cite{jordan1993algebraic}. In this context, the Jordan algebras represent algebras of observables, because unlike the usual matrix product, the commutative but non-associative Jordan product of two Hermitian operators is again Hermitian. In particular, the classification of Jordan algebras identified the single exceptional Jordan algebra~$J_3(\bb{O})$, consisting of $3\times 3$-matrices with (off-diagonal) entries in~$\bb{O}$. 

Not long after the discovery of quarks, G\"uynadin and G\"ursey combined these earlier ideas and utilized~$\bb{O}$ to describe the~$\SU(3)$ color symmetry of quarks~\cite{gunaydin1973quark}. Such an octonion-based description of quarks is motivated by the observation that~$\SU(3)$ corresponds to a subgroup of the octonion automorphism group~$G_2$ that leaves invariant one of the octonion imaginary unit vectors.

This early work relating~$\bb{O}$ to quarks was extended upon by Dixon who demonstrated that the division algebras~$\bb{C}$, $\bb{H}$ and~$\bb{O}$ provide the algebraic underpinnings of modern particle theory. Specifically, many of the mathematical features of the standard model, including its gauge symmetries to which a single generation of fermions and leptons are subject, together with the correct multiplets into which they fall, are inherent in the 64 real dimensional composition algebra~$\bb{T}=\mathbb{R}\otimes\mathbb{C}\otimes\mathbb{H}\otimes\mathbb{O}$,
\cite{dixon1990derivation,dixon19991,dixon1999algebraic,dixon2004division,dixon2010division,dixon2013division}.

Conventionally, spinors are considered as column vectors with entries in~$\bb{R}$ or~$\bb{C}$. An example being Dirac spinors corresponding to wave functions in $4\times 1$-column vectors over~$\bb{C}$, acted upon by the Dirac algebra (a complex Clifford algebra). Dixon's underlying idea is to generalize this notion and consider tensor products of division algebras (for example~$\bb{T}$ defined above) themselves as spinors.

Such spinors are acted upon from the left or right by the algebra itself, with each possible action corresponding to a linear map. The composition of all such adjoint actions generate the \textit{associative multiplication algebra}, a Clifford algebra, which acts on the spinor space (corresponding to a tensor product of division algebras).

In such a construction therefore, the spinor space itself exhibits algebraic structure. It is this nontrivial algebraic structure of the spinor components that provides the algebraic source of the internal symmetries needed to replicate the structure of the standard model. In particular, since~$\bb{T}$ itself is not a division algebra, it admits a nontrivial decomposition of its identity into mutually orthogonal primitive idempotents, corresponding to projectors. These projectors simultaneously decompose the spinor space into mutually orthogonal subspaces, as well as the associative multiplication algebra into subalgebras whose actions preserve these subspaces. 

For the case of the spinor~$\bb{T}$, the identity can be resolved into four orthogonal projectors, each composed of other projectors associated with~$\SU(2)$ isospin and~$\SU(3)$ color. The spinor~$\bb{T}$ is then found to transform with respect to these Lie groups as~$\SU(2)$ doublets and~$\SU(3)$ singlets and triplets. Additionally, the quantum numbers associated with hypercharge, isospin and color can be mathematically assigned to the invariant subspaces of~$\bb{T}$. That is, the gauge symmetries and multiplets of spinors encountered in the standard model are both part of the same mathematical structure arising from the spinor~$\bb{T}$, which then transforms as~$\U(1)\times \SU(2)\times \SU(3)$, with each invariant subspace corresponding to a Pauli spinor. Dirac spinors can subsequently be produced by doubling the spinor space to~$\bb{T}^2$, that is $2\times 1$-column vectors over~$\bb{T}$.

In the models of G\"uynadin and G\"ursey~\cite{gunaydin1973quark} and Dixon~\cite{dixon2013division}, 
the bosons live in the associative multiplication algebra, whereas the fermions reside in the spinors on which this algebra acts. Furey instead describes both bosons and fermions within the associative multiplication algebra by considering the minimal left ideals of the algebra as the spinors~\cite{furey2016standard,furey2018demonstration,furey231,furey232,furey233}. The representation of spinors as minimal ideals of Clifford algebras dates back to the 1930s~\cite{juvet1930operateurs,sauter1930losung} and 1940s~\cite{riesz2013clifford}. The fermions in such a construction then correspond to the basis states of the minimal left ideals of the associative multiplication algebra, constructed via a standard procedure based on a Witt decomposition proposed by Ablamowicz~\cite{ablamowicz1995construction}, and the gauge symmetries are then the unitary symmetries that preserve these ideals.  

Since the predominant focus of the present paper is connecting causal fermion systems to the octonions, 
in what follows we shall disregard the factor of~$\bb{H}$ in~$\bb{T}$. Instead, we restrict attention to
the~$\bb{C}\otimes\bb{O}$-component of the algebra~$\bb{T}$.

\subsection{The Normed Division Algebras}

A division algebra is an algebra over a field where division is always well-defined, except by zero. A normed division algebra has the additional property that it is also a normed vector space, with the norm defined in terms of a conjugate. A well-known result by Hurwitz~\cite{hurwitz1898ueber} states that there exist only four normed division algebras (over the field of real numbers): $\bb{R}$, $\bb{C}$, $\bb{H}$, $\bb{O}$, being of dimensions one, two, four and eight, respectively. In going to higher-dimensional algebras, successive algebraic properties are lost: $\bb{R}$ is self-conjugate, commutative and associative, $\bb{C}$ is commutative and associative (but no longer self-conjugate), $\bb{H}$ is associative but no longer commutative, and finally~$\bb{O}$ is neither commutative nor associative (but alternative).

The octonions $\bb{O}$ are the largest division algebra, of dimension eight. Its orthonormal basis is comprised of the seven imaginary units~$e_1,...,e_7$ along with the unit~$1=e_0$.
A general octonion~$x$ may then be written as
\[ x=x_0\, e_0+x_1\, e_1+...+x_7\, e_7 \qquad \text{with} \qquad x_0,...,x_7\in\bb{R} \:. \]
The octonionic {\em{conjugate}}~$\overline{x}$ is defined as~$\overline{x}=x_0e_0-x_1e_1-...-x_7e_7$. The
{\em{norm}} of an octonion~$x$ is subsequently defined by~$\vert\vert x\vert\vert^2=x\overline{x}=\overline{x}x$,
and the {\em{inverse}} of $x$ is given by~$x^{-1}=\overline{x}/\vert\vert x\vert\vert^2$. 

\begin{figure}[h!]
\centering
\includegraphics[scale=0.08]{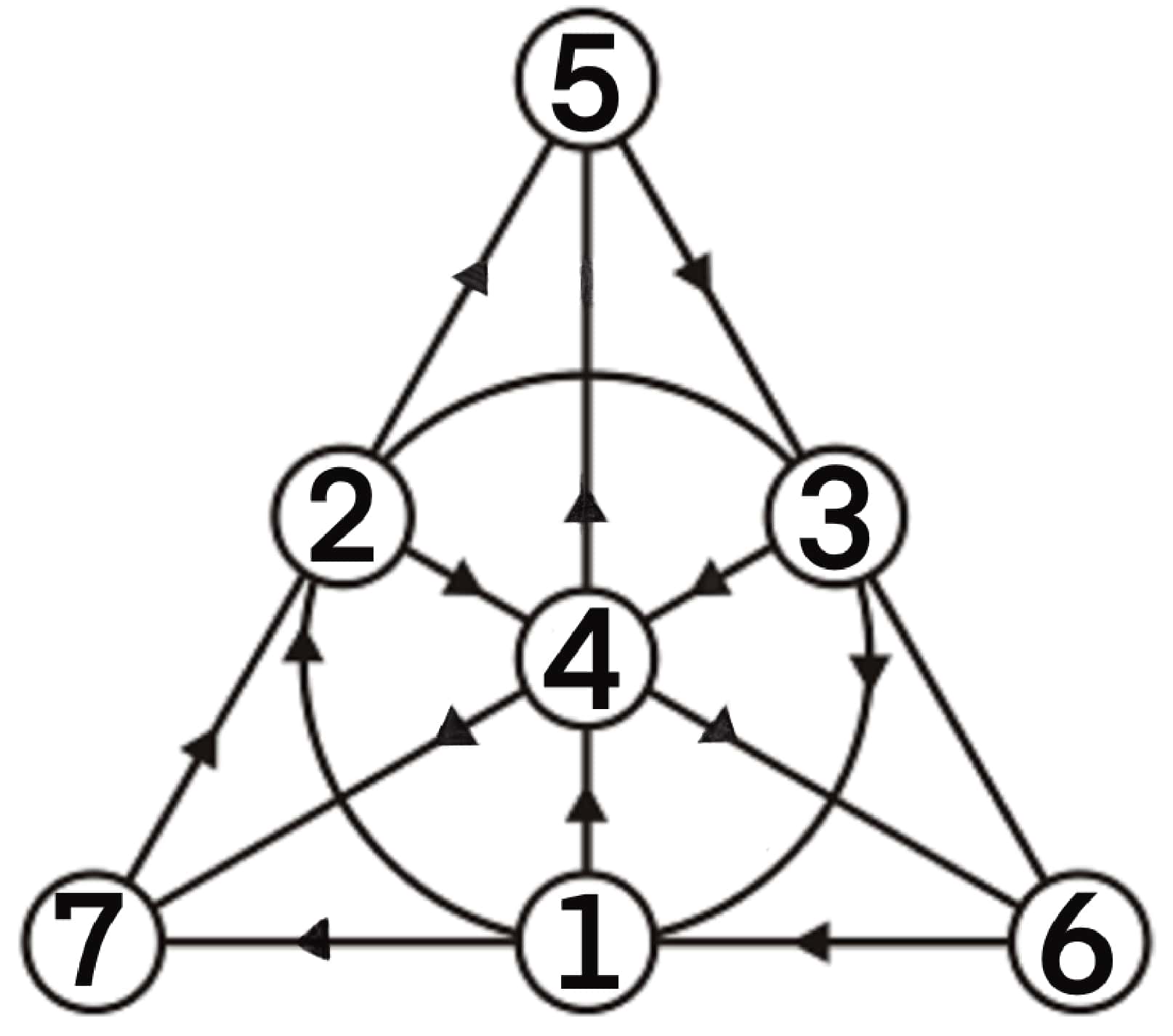}
\caption{The Fano plane, encoding the multiplicative structure of our octonions, where~$a\equiv e_a,\; a=1,...,7$. Note that each line is cyclic, representing a quaternionic triple.}
\label{fanoplane}
\end{figure}

There are many different multiplication tables for octonions, with different authors using different multiplication rules. Here we will follow the multiplication table used in~\cite{lohmus1994nonassociative}, which is the multiplication table most naturally obtained via the Cayley-Dickson process. That is the multiplication of~$e_0,e_1,e_2,e_3$ generates~$\bb{H}$ and the multiplication of~$e_0,e_1$ generates~$\bb{C}$ (and of course~$e_0$ generates~$\bb{R}$ trivially). One way to represent octonion multiplication is in terms of the {\em{Fano plane}}; see Figure~\ref{fanoplane}. 
\vspace{5pt}

\noindent
The Fano plane representing the multiplication of octonion basis elements can be summarized as:
\[ e_ie_j= \left\{ \begin{array}{cl}
        e_j&\textrm{if}\;i=0\\
        e_i&\textrm{if}\;j=0\\
        -\delta_{ij}+\epsilon_{ijk}e_k&\textrm{otherwise}\:,
    \end{array} \right. \]
where~$\epsilon_{ijk}$ is the completely antisymmetric tensor with value 1 when~$ijk = 123$, $145$, $176$,
$246$, $257$, $347$ or~$365$.

Each projective line in the Fano plane corresponds (together with the identity $e_0$) to a quaternion ($\bb{H}$) subalgebra; there are seven such subalgebras. All the imaginary octonion units anti-commute under multiplication. Unlike for the smaller division algebras, the multiplication of elements not belonging to the same $\bb{H}$ subalgebra is {\em{non-associative}}. For example,
\[ e_4(e_7e_6) = -e_5 \neq e_5 = (e_4e_7)e_6 \:. \]
Octonion multiplication however is {\em{alternative}}, meaning that
\[ x(xy) = (xx)y \quad\text{and} \quad y(xx) = (yx)x \qquad \text{for all~$x,y\in\bb{O}$}\:. \]

By singling out one imaginary octonion basis element, an octonion can be written as
\begin{eqnarray} \label{2.4}
x=(x_0+x_4e_4)e_0+(x_1-x_5e_4)e_1+(x_2-x_6e_4)e_2+(x_3-x_7e_4)e_3 \:.
\end{eqnarray}
This singling out of one of the imaginary basis elements of the octonions is equivalent to choosing a privileged subalgebra of~$\C \otimes \bb{O}$, and induces a splitting of~$\bb{O}$ as~$\bb{C}\oplus\bb{C}^3$. Therefore, as vector spaces, the octonions~$\bb{O}$ are therefore isomorphic to~$\bb{C}^4$. this splitting of~$\bb{O}$ as~$\bb{C}\oplus\bb{C}^3$ has been associated with the lepton-quark splitting~\cite{manogue1999dimensional}. The space of internal states of a quark then corresponds to the three complex dimensional space~$\bb{C}^3$, whereas the internal space of a lepton is~$\bb{C}$. 

The automorphism group of~$\bb{O}$ is the~$14$-dimensional exceptional Lie group~$G_2$. This exceptional group contains~$\SU(3)$ as one of its maximal subgroups, corresponding to the stabilizer subgroup of one of the octonion imaginary units or, equivalently, the subgroup of~$\text{Aut}(\bb{O})$ that preserves the representation of~$\bb{O}$ as the complex space~$\bb{C}\oplus \bb{C}^3$.

On account of being a division algebra, the octonions do not contain any projectors. One way to include projectors it to instead consider the complex octonions~$\bb{C}\otimes \bb{O}$\footnote{Another option is to consider the split octonions algebra, which is not a normed division algebra, and does contain projectors. The split octonions differ from the octonions in that its quadratic form has a split signature~$(4,4)$ whereas the octonions quadratic form has a positive-definite signature~$(8,0)$.}, and this is one reason why in octonionic models it is inevitably the complex octonion algebra $\bb{C}\otimes\bb{O}$ that is considered, and not just $\bb{O}$. The algebra $\bb{C}\otimes\bb{O}$ is no longer a division algebra (even though the separate factors are), but~$\bb{C}\otimes \bb{O}$  remains alternative.  Two specific projectors can be introduced in~$\bb{C}\otimes \bb{O}$ by singling out one of the octonion basis elements, say~$e_4$. The the two mutually annihilating projectors are
\[ \rho_{\pm}=\frac{1}{2}(1 \pm ie_4) \:, \]
satisfying the relations
\[ \rho_-+\rho_+=1,\quad \rho_-^2=\rho_-,\quad \rho_+^2=\rho_+,\quad \rho_-\rho_+=\rho_+\rho_-=0 \:. \]

\subsection{The Associative Multiplication Algebras of~$\bb{O}$}

The non-associa\-ti\-vity of~$\bb{O}$ means that the algebra is not representable as a matrix algebra (with the standard matrix product). The algebra generated from the composition of left (and/or right) actions of~$\bb{O}$ on itself, however, is associative, since each such left (right) action corresponds to a linear operator (endomorphism). 
Suppose that~$a,x$ are elements of some algebra~$\bb{A}$, and define:
\[ L_a[x]:=ax,\quad R_a[x]:=xa,\quad \forall a,x\in  \bb{A} \:. \]
It is clear that~$L_a$ and~$R_a$ denote linear operator corresponding to left and right multiplication of~$x\in \bb{A}$ by~$a\in \bb{A}$, respectively. These maps send the element~$x\in\bb{A}$ to some other element~$y\in\bb{A}$, and thus~$L_a$ and~$R_a\in \text{End}(\bb{A)}$. The mappings~$a\mapsto L_a$ and~$a\mapsto R_a$ do not correspond to algebra homomorphisms. Instead, via composition each generates an associative algebra referred to as the \textit{associative multiplication algebra}.

Note in particular, that for a non-associative algebras such as~$\bb{O}$, 
\begin{align*}
    L_aL_b[x] &=a(bx) \quad \neq \quad L_{ab}[x]=(ab)x,\\
    R_aR_b[x] &=(xb)a \quad \neq \quad R_{ab}[x]=x(ab),\quad\forall a,b,x\in\bb{O}.
\end{align*}
We denoted by~$\bb{O}_L$ and~$\bb{O}_R$, the left and right associative multiplication  algebras of octonions respectively, which we will henceforth refer to simply as the {\em{left (right) multiplication algebra}}\footnote{We remark that the left multiplication algebra is also referred to as the
{\em{octonionic chain algebra}}~$\C \otimes \overleftarrow{\mathbb{O}}$; see
for example~\cite{furey2016standard}.}
These linear maps preserve the quadratic relations~$\langle x,y\rangle=\frac{1}{2}(x\overline{y}+y\overline{x})$,
where~$x,y\in \bb{O}$
\[ L_xL_{\overline{y}}1+L_yL_{\overline{x}}1=2\langle x,y \rangle 1=R_xR_{\overline{y}}1+R_yR_{\overline{x}}1 \:. \]

Since~$L_a$ ($R_a$) correspond to linear operators, they can be represented as $8\times 8$ real matrices acting on the vector space~$\bb{O}$ written as a column vector.

Within~$\bb{O}_L$, one finds the following identities~\cite{dixon2013division,furey2016standard}:
\begin{align*}
    L_{e_1}L_{e_2}L_{e_3}L_{e_4}L_{e_5}L_{e_6}x &= L_{e_7}x,\\
    ...L_{e_b}L_{e_a}L_{e_a}L_{e_c}....x &= -...L_{e_b}L_{e_c}....x,\\
    ...L_{e_a}L_{e_b}...x &= -...L_{e_b}L_{e_a}...,
\end{align*}
where~$a,b,c=1,...,7$.

The non-associativity of~$\bb{O}$ means that~$\bb{O}_L$ ($\bb{O}_R$) contains genuinely new maps which are not captured by~$\bb{O}$ itself. For example, $e_3(e_4(e_6+e_2))\neq y(e_6+e_2)$ for any~$y\in\bb{O}$. 
There are a total of 64 distinct left-acting real linear maps from~$\bb{O}$ to itself, and these (due to the given identities above) provide a faithful representation of the real Clifford algebra~$C\ell(0,6)$. Furthermore, any left (right) adjoint action can always be rewritten as a right (left) adjoint action, so that~$\bb{O}_L=\bb{O}_R$~\cite{dixon2013division}.

For the complex octonions, $\bb{C}\otimes\bb{O}$, the left (right) multiplication algebra is likewise complexified and becomes the complex Clifford algebra~$\bb{C}\ell(6)$, having complex dimension~$64$. 

\subsection{Connection to the Standard Model} \label{sec:octsm}
In order to highlight the connection of the octonions to the standard mode, we now provide an explicit example of how the associative left multiplication algebra $\bb{C}\ell(6)$ generated from the adjoint actions of $\bb{C}\otimes\bb{O}$ can be used to represent one generation of standard model fermions with $\SU(3)\times \U(1)$ unbroken gauge symmetries.

\subsubsection{One Generation of Electrocolor States from~$\bb{C}\otimes\bb{O}$} \label{Furey}
Let~$L_{e_1},...,L_{e_6}$ be a generating basis (over~$\bb{C}$) for~$\bb{C}\ell(6)$ associated with the left associative multiplication algebra of~$\bb{C}\otimes\bb{O}$, satisfying the relations~$L_{e_i}^2=-1$ and~$L_{e_i}L_{e_j}=-L_{e_j}L_{e_i}$. Define a Witt basis for~$\bb{C}\ell(6)$ as follows,
\begin{align*}
\alpha_1^\dagger &:= \frac{1}{2}(L_{e_1}+iL_{e_5}),& \alpha_2^\dagger &:= \frac{1}{2}(L_{e_2}+iL_{e_6}),& \alpha_3^\dagger &:= \frac{1}{2}(L_{e_3}+iL_{e_7}) \\
\alpha_1 &:= \frac{1}{2}(-L_{e_1}+iL_{e_5}), & \alpha_2 &:= \frac{1}{2}(-L_{e_2}+iL_{e_6}),& \alpha_3 &:= \frac{1}{2}(-L_{e_3}+iL_{e_7}) \:.
\end{align*}
Here~${}^{\dagger}$ corresponds to the composition of complex conjugation and octonion conjugation. This new basis satisfies the anti-commutation relations
\[ \{\alpha_i,\alpha_j\}\, x = \{\alpha_i^{\dagger},\alpha_j^{\dagger}\}\, x=0,
\quad\{\alpha_i,\alpha_j^{\dagger}\}\, x=\delta_{ij}x,\qquad \text{for all~$x\in\bb{C}\otimes\bb{O}$} \:. \]
Both~$\{\alpha_i\}$ and~$\{\alpha_i^{\dagger}\}$ correspond to bases for the maximally  isotropic subspaces~$\chi$ and~$\chi^{\dagger}$ respectively, because
\[ \{\alpha_i,\alpha_j\}=0 \quad \forall \alpha_i,\alpha_j\in \chi
\qquad \text{and} \qquad \{\alpha_i^{\dagger},\alpha_j^{\dagger}\}=0 \quad \forall \alpha_i^{\dagger},\alpha_j^{\dagger}\in\chi^{\dagger} \:. \]
Moreover, they generate the exterior algebras~$\bigwedge \chi$ and~$\bigwedge\chi^{\dagger}$ via the product of the Clifford algebra.

From this Witt basis it is possible to construct two minimal left ideals of the algebra~$\bb{C}\ell(6)$, following the procedure in~\cite{ablamowicz1995construction} (see~\cite{furey2016standard} for a detailed construction):
\[ \bb{C}\ell(6)\,\omega\omega^{\dagger}=\bigwedge\chi^{\dagger}\omega\omega^{\dagger},\qquad \bb{C}\ell(6)\,\omega^{\dagger}\omega=\bigwedge\chi\omega^{\dagger}\omega \:, \]
where~$\omega = \alpha_1\alpha_2\alpha_3$ and~$\omega^\dagger = \alpha_3^\dagger\alpha_2^\dagger\alpha_1^\dagger$ are nilpotents, but~$\omega\omega^{\dagger}$ and~$\omega^{\dagger}\omega$ are primitive idempotents. Each minimal ideal has complex dimension eight. Explicitly,
\begin{equation} \notag
 \begin{split}
     S^u=&{\nu}\omega\omega^\dagger + \\  \overline{d^r}{\alpha_1^\dagger}\omega\omega^\dagger + &\overline{d^g}{\alpha_2^\dagger}\omega\omega^\dagger + \overline{d^b}{\alpha_3^\dagger}\omega\omega^\dagger + \\ u^r{\alpha_3^\dagger}{\alpha_2^\dagger}\omega\omega^\dagger + &u^g{\alpha_1^\dagger}{\alpha_3^\dagger}\omega\omega^\dagger + u^b{\alpha_2^\dagger}{\alpha_1^\dagger}\omega\omega^\dagger + \\ &e^+{\alpha_3^\dagger}{\alpha_2^\dagger}{\alpha_1^\dagger}\omega\omega^\dagger
\end{split}
\qquad 
\begin{split}
     S^d=&{\overline{\nu}}\omega^\dagger\omega + \\ {d}^r{\alpha_1}\omega^\dagger\omega + &{d}^g{\alpha_2}\omega^\dagger\omega +  {d}^b{\alpha_3}\omega^\dagger\omega + \\ \overline{u^r}{\alpha_3}{\alpha_2}\omega^\dagger\omega + &\overline{u^g}{\alpha_1}{\alpha_3}\omega^\dagger\omega + \overline{u^b}{\alpha_2}{\alpha_1}\omega^\dagger\omega + \\ &e^-{\alpha_3}{\alpha_2}{\alpha_1}\omega^\dagger\omega
\end{split}
\end{equation}
where the suggestively labeled coefficients are elements of~$\bb{C}$. 

The subgroup of~$\Spin(6)$ generated from the grade two elements of~$\bb{C}\ell(6)$ that preserves the Witt basis, and hence the minimal left ideals, is the unitary symmetry~$\U(3)=\SU(3)\times \U(1)$. The generators of this symmetry, written in terms of the Witt basis elements, are:
    \begin{align} \notag
     \Lambda_1 &= -\alpha_2^\dagger\alpha_1 - \alpha_1^\dagger\alpha_2 \notag \hspace{2em}
     \Lambda_2 = i\alpha_2^\dagger\alpha_1 -i\alpha_1^\dagger\alpha_2 \notag \hspace{2em}
     \Lambda_3 = \alpha_2^\dagger\alpha_2 - \alpha_1^\dagger\alpha_1 \notag\\ 
     \Lambda_4 &= -\alpha_1^\dagger\alpha_3 - \alpha_3^\dagger\alpha_1 \notag \hspace{2em}
     \Lambda_5 = -i\alpha_1^\dagger\alpha_3 + i\alpha_3^\dagger\alpha_1 \notag \hspace{2em}
     \Lambda_6 = -\alpha_3^\dagger\alpha_2 - \alpha_2^\dagger\alpha_3 \notag\\
     \Lambda_7 &= i\alpha_3^\dagger\alpha_2 - i\alpha_2^\dagger\alpha_3 \notag \hspace{1em}
     \Lambda_8 = -\frac{1}{\sqrt{3}}(\alpha_1^\dagger\alpha_1 + \alpha_2^\dagger\alpha_2 - 2\alpha_3^\dagger\alpha_3), \notag
 \end{align}
     \begin{equation} \notag
     Q = \frac{1}{3}(\alpha_1^\dagger\alpha_1 + \alpha_2^\dagger\alpha_2 + \alpha_3^\dagger\alpha_3).
 \end{equation}
The basis states of minimal ideals are then found to transform as~$1\oplus 3\oplus\overline{3}\oplus1$ under~$\SU(3)$, which can therefore be associated with the color symmetry~$\SU(3)_C$, justifying the choice of coefficients. The~$\U(1)$ generator~$Q$ (together with~$-Q^*$), related to the number operator~$Q=N/3$, on the other hand, gives the correct electric charge for each state. The ideal~$S^u$ contains the isospin up states, whereas the~$S^d$ contains the isospin down states.

Two minimal ideal of ~$\bb{C}\ell(6)$ arising as the associative left multiplication algebra of ~$\bb{C}\otimes\bb{O}$ therefore provides an elegant representation of one generation of standard model electrocolor states. In this construction, the dimension of the minimal ideals (eight) corresponds to the number of distinct particle states, whereas the gauge symmetries are obtained as the subgroup of~$\Spin(6)$ that preserves the Witt basis (and subsequently the minimal left ideals).

\section{Causal Fermion Systems} \label{seccfsintro}
We now give a self-contained introduction to the basic structures of a causal fermion system.
We keep the explanations short, noting that detailed and elementary introductions
to the physical and mathematical concepts can be found in the review papers~\cite{dice2014, review}
and the comparison paper~\cite[Section~2.1]{mmt-cfs}.
We also refer the interested reader to the textbooks~\cite{cfs, intro} and the website~\cite{cfsweblink}.

\subsection{Causal Fermion Systems and the Reduced Causal Action Principle}
We begin with the general definition of a causal fermion system.
\begin{Def} \label{defcfs} (causal fermion systems) {\em{ 
Given a separable complex Hilbert space~$\H$ with scalar product~$\la .|. \ra_\H$
and a parameter~$n \in \N$ (the {\em{``spin dimension''}}), we let~$\F \subset \Lin(\H)$ be the set of all
symmetric operators on~$\H$ of finite rank, which (counting multiplicities) have
at most~$n$ positive and at most~$n$ negative eigenvalues. On~$\F$ we are given
a positive measure~$\rho$ (defined on a $\sigma$-algebra of subsets of~$\F$).
We refer to~$(\H, \F, \rho)$ as a {\em{causal fermion system}}.
}}
\end{Def} \noindent

A causal fermion system describes a spacetime together
with all structures and objects therein.
In order to single out the physically admissible
causal fermion systems, one must formulate physical equations. To this end, we impose that
the measure~$\rho$ should be a minimizer of the causal action principle,
which we now introduce. For brevity of the presentation, we only consider the
{\em{reduced causal action principle}} where the so-called boundedness constraint has been
incorporated by a Lagrange multiplier term. This simplification is no loss of generality, because
the resulting Euler-Lagrange (EL) equations are the same as for the non-reduced action principle
as introduced for example in~\cite[Section~\S1.1.1]{cfs}.

For any~$x, y \in \F$, the product~$x y$ is an operator of rank at most~$2n$. 
However, in general it is no longer a symmetric operator because~$(xy)^* = yx$,
and this is different from~$xy$ unless~$x$ and~$y$ commute.
As a consequence, the eigenvalues of the operator~$xy$ are in general complex.
We denote the rank of~$xy$ by~$k \leq 2n$. Counting algebraic multiplicities, we choose~$\lambda^{xy}_1, \ldots, \lambda^{xy}_{k} \in \C$ as all the non-zero eigenvalues and set~$\lambda^{xy}_{k+1}, \ldots, \lambda^{xy}_{2n}=0$.
Given a parameter~$\kappa>0$ (which will be kept fixed throughout this paper),
we introduce the $\kappa$-Lagrangian and the causal action by
\begin{align}
\text{\em{$\kappa$-Lagrangian:}} && \L(x,y) &= 
\frac{1}{4n} \sum_{i,j=1}^{2n} \Big( \big|\lambda^{xy}_i \big|
- \big|\lambda^{xy}_j \big| \Big)^2 + \kappa\: \bigg( \sum_{j=1}^{2n} \big|\lambda^{xy}_j \big| \bigg)^2 \label{Lagrange} \\
\text{\em{causal action:}} && \Sact(\rho) &= \iint_{\F \times \F} \L(x,y)\: d\rho(x)\, d\rho(y) \:. \label{Sdef}
\end{align}
The {\em{reduced causal action principle}} is to minimize~$\Sact$ by varying the measure~$\rho$
under the following constraints,
\begin{align}
\text{\em{volume constraint:}} && \rho(\F) = 1 \quad\;\; \label{volconstraint} \\
\text{\em{trace constraint:}} && \int_\F \tr(x)\: d\rho(x) = 1 \:. \label{trconstraint}
\end{align}
This variational principle is mathematically well-posed if~$\H$ is finite-dimensional.
For a review of the existence theory and the analysis of general properties of minimizing measures
we refer to~\cite[Chapter~12]{intro}.

\subsection{Spacetime Structures}
A causal fermion system describes a spacetime together with all structures therein.
The structures are the causal and metric structures, spinors
and interacting fields (for details see~\cite[Chapter~1]{cfs}).
All these spacetime structures are {\em{inherent}} in the sense that they use information
already encoded in the causal fermion system.
We now give a brief outline of the spacetime structures, with a focus on those structures
which will be needed later on.

Let~$\rho$ be a {\em{minimizing}} measure. {\em{Spacetime}}
is defined as the support of this measure,
\[ 
M := \supp \rho \;\subset\; \F \:, \]
where on~$M$ we consider the {\em{topology}} induced by~$\F$ (generated by the operator norm
on~$\Lin(\H)$).
Thus the spacetime points are symmetric linear operators on~$\H$.
The restriction of the measure~$\rho|_M$ gives a volume measure on spacetime.

We begin with the following notion of causality:
\begin{Def} (causal structure) \label{def2} 
{\em{ For any~$x, y \in \F$, we again denote the non-trivial ei\-gen\-values of the operator product~$xy$
(again counting algebraic multiplicities) by~$\lambda^{xy}_1, \ldots, \lambda^{xy}_{2n}$.
The points~$x$ and~$y$ are
called {\em{spacelike}} separated if all the~$\lambda^{xy}_j$ have the same absolute value.
They are said to be {\em{timelike}} separated if the~$\lambda^{xy}_j$ are all real and do not all 
have the same absolute value.
In all other cases (i.e.\ if the~$\lambda^{xy}_j$ are not all real and do not all 
have the same absolute value),
the points~$x$ and~$y$ are said to be {\em{lightlike}} separated. }}
\end{Def} \noindent
Restricting the causal structure of~$\F$ to~$M$, we get causal relations in spacetime.

The Lagrangian~\eqref{Lagrange} is compatible with the above notion of causality in the
following sense.
Suppose that two points~$x, y \in M$ are spacelike separated.
Then the eigenvalues~$\lambda^{xy}_i$ all have the same absolute value.
As a consequence, the Lagrangian~\eqref{Lagrange} vanishes. Thus pairs of points with spacelike
separation do not enter the action. This can be seen in analogy to the usual notion of causality where
points with spacelike separation cannot influence each other.
This is the reason for the notion ``causal'' in {\em{causal}} fermion system
and {\em{causal}} action principle.

Next, a causal fermion system encodes {\em{spinorial wave functions}}.
To this end, for every spacetime point~$x \in M$ we define the {\em{spin space}}~$S_x$ by~$S_x = x(\H)$;
it is a subspace of~$\H$ of dimension at most~$2n$.
It is endowed with the {\em{spin inner product}}~$\Sl .|. \Sr_x$ defined by
\beq \label{ssp}
\Sl u | v \Sr_x = -\la u | x v \ra_\H \qquad \text{(for all~$u,v \in S_x$)}\:.
\eeq
A {\em{wave function}}~$\psi$ is defined as a function
which to every~$x \in M$ associates a vector of the corresponding spin space,
\[ 
\psi \::\: M \rightarrow \H \qquad \text{with} \qquad \psi(x) \in S_x \quad \text{for all~$x \in M$}\:. \]
It is an important observation that every vector~$u \in \H$ of the Hilbert space gives rise to a distinguished
wave function. In order to obtain this wave function, denoted by~$\psi^u$, we simply project the vector~$u$
to the corresponding spin spaces,
\[ 
\psi^u \::\: M \rightarrow \H\:,\qquad \psi^u(x) = \pi_x u \in S_x \:. \]
We refer to~$\psi^u$ as the {\em{physical wave function}} of~$u \in \H$.
Another object which will be important in what follows is the 
{\em{kernel of the fermionic projector}} defined by
\beq \label{Pxyabs}
P(x,y) = \pi_x \,y|_{S_y} \::\: S_y \rightarrow S_x \:,
\eeq
where~$\pi_x : \H \rightarrow S_x$ is the orthogonal projection to the spin space~$S_x$.
The kernel of the fermionic projector can be expressed in terms of the physical wave functions.
Indeed, choosing an orthonormal basis~$(b_i)$ of~$\H$
(for details see~\cite[Chapter~1]{cfs} or~\cite[Section~5.6]{intro})
\beq \label{Pxyphysical}
P(x,y) \, \phi = -\sum_i \psi^{b_i}(x) \: \Sl \psi^{b_i}(y) |\; \phi \Sr_y \:.
\eeq

The kernel of the fermionic projector plays a central role in the theory of causal fermion systems
for two reasons:
\bitem
\item[(i)] The causal action principle can be formulated in terms of~$P(x,y)$.
In order to see this, we note that the trace in the trace constraint~\eqref{trconstraint}
can be written as~$\tr(x) = \Tr_{S_x}(P(x,x))$. Moreover, the eigenvalues~$\lambda^{xy}_j$ of the operator~$xy$
(which appear in~\eqref{Lagrange} and~\eqref{Sdef}) coincide with the eigenvalues
of the so-called closed chain~$A_{xy} := P(x,y)\, P(y,x) : S_x \rightarrow S_x$.
Details can be found in~\cite[Section~1.1]{cfs} and~\cite[Section~5.6]{intro}.
\item[(ii)] According to~\eqref{Pxyabs}, the kernel of the fermionic projector~$P(x,y)$ is a mapping from one space~$S_y$
to another spin space~$S_x$. In this way, it gives relations between the spacetime points.
As will be explained in more detail in Section~\ref{secqg}, these relations give rise to {\em{geometric structures}}
like connection and curvature.
\eitem
Finally, additional structures like the {\em{metric}} and {\em{bosonic fields}} arise when
describing the dynamics of the causal fermion system, as will be outlined in the next sections.

\subsection{Dynamical Equations in Spacetime} \label{secdynamics}
A minimizer of the causal action principle satisfies the following {\em{Euler-Lagrange (EL) equations}}.
For a suitable value of the parameter~$\s>0$,
the function~$\ell : \F \rightarrow \R_0^+$ defined by
\beq \label{elldef}
\ell(x) := \int_M \L_\kappa(x,y)\: d\rho(y) - \s
\eeq
is minimal and vanishes on spacetime~$M:= \supp \rho$,
\beq \label{EL}
\ell|_M \equiv \inf_\F \ell = 0 \:.
\eeq
Here the parameter~$\s \geq 0$ in~\eqref{elldef} is the Lagrange parameter
corresponding to the volume constraint. For the derivation and further details we refer to~\cite[Section~2]{jet}
or~\cite[Chapter~7]{intro}.

The EL equations~\eqref{EL} describe the dynamics of the causal fermion systems.
This can be done abstractly on the linearized level by working with the so-called {\em{linearized fields}}
which satisfy the {\em{linearized field equations}}.
Under general assumptions, the Cauchy problem for the linearized field equations is well-posed~\cite{linhyp},
giving rise to a causal propagation with finite propagation speed.
Starting from the linearized dynamics, the EL equations can be treated perturbatively~\cite{perturb}.

The abstract ``fields'' in the linearized field equations can be associated in the concrete applications with
classical or quantum fields in the usual sense. We proceed with a brief outline of these constructions
and the obtained results.

\subsection{Connection to the Standard Model} \label{seccfssm}
In~\cite[Chapter~5]{cfs} it is shown that in a specific limiting case, referred to as the {\em{continuum limit}},
the causal action principle gives rise to the interactions of the standard model and general relativity (GR)
in terms of an interaction of Dirac particles and anti-particles with classical bosonic fields
(the connection to bosonic quantum fields will be made in Section~\ref{secqft} below).
The analysis of the continuum limit is a systematic procedure for evaluating the EL equations for causal fermion systems constructed in Minkowski space on the level of classical bosonic fields
coupled to second-quantized fermionic fields.

The input is to prescribe the configuration of the fermions in the vacuum.
To this end, one needs to specify the kernel of the fermionic projector of the vacuum~$P(x,y)$
as given abstractly by~\eqref{Pxyabs}.
In view of~\eqref{Pxyphysical}, we need to specify all the physical wave functions. 
Choosing them as Dirac wave functions in Minkowski space, the
kernel of the fermionic projector~$P(x,y)$ becomes a spinorial bi-distribution in Minkowski space.
More specifically, in order to describe the vacuum, one chooses the physical wave functions
as all the negative energy solutions of the Dirac equation (the {\em{Dirac sea}}; for the physical
background see~\cite[Section~1.5]{intro} or~\cite[Chapter~1]{cfs}).
Describing different types of particles by one Dirac sea each, we are led to
building up the vacuum of many Dirac seas.
Indeed, as described in~\cite[Chapter~5]{cfs}, the vacuum is described by the fermionic projector
\beq \label{Pvac}
P(x,y) = P^N(x,y) \oplus P^C(x,y) \:,
\eeq
where the {\em{charged component}}~$P^C$ is formed as the direct sum of seven
identical sectors, each consisting of a sum of three Dirac seas,
\beq \label{PC}
P^C(x,y) = \bigoplus_{a=1}^7 \sum_{\beta=1}^3 P^\text{vac}_{m_\beta}(x,y) \:,
\eeq
where~$m_\beta$ are the masses of the fermions and~$P^\text{vac}_m$ is the distribution
\beq \label{Psea}
P^\text{vac}_m(x,y) = \int \frac{d^4k}{(2 \pi)^4}\: (\slashed{k}+m)\: \delta(k^2-m^2)\: \Theta(-k^0)\: e^{-ik(x-y)}\:.
\eeq
We point out that, in view of the Dirac matrices, this distribution is a linear operator on spinors,
mapping a spinor at the spacetime point~$y$ to a spinor at~$x$.

Likewise, for the {\em{neutrino sector}}~$P^N$ we choose the ansatz of potentially massive neutrinos
\beq \label{massneutrino}
P^N(x,y) = X \sum_{\beta=1}^3 P^\text{vac}_{\tilde{m}_\beta}(x,y) \:.
\eeq
The neutrino masses~$\tilde{m}_\beta \geq 0$ will in general be different from
the masses~$m_\beta$ in the charged sectors\footnote{In order to avoid confusion, we note that
these masses should be considered as the ``bare masses,'' i.e.\ the masses on the Planck energy.
Due to the self-interaction, these masses are different from the physical masses to be measured
in experiments. In the ansatz~\eqref{PC}, within each generation,
the quarks and the charged leptons have the same bare mass.}.
At most two of the neutrino masses could be zero.

We next introduce an {\em{ultraviolet regularization}} on the length scale~$\varepsilon$.
The regularized vacuum fermionic projector is denoted by~$P^\varepsilon$.
We again use the formalism of the continuum limit as developed in~\cite[Chapter~4]{pfp}
(see also~\cite[Section~3.5]{cfs}). An important physical input is that, in the {\em{neutrino sector}},
we work with a {\em{regularization}} which {\em{breaks the chiral symmetry}}
(for technical details see~\cite[\S4.2.5]{cfs}).
We point out that this breaking of the chiral symmetry is built in by hand.
It is part of the physical input needed to specify the causal fermion system of the vacuum.

Taking the so-regularized kernel of the vacuum fermionic projector as the starting point,
in the continuum limit one studies the EL equations of the causal action principle
for systems which are obtained from the vacuum by introducing an interaction via
classical bosonic potentials. Before giving an outline on how the continuum limit analysis works,
we briefly mention the results as derived in~\cite[Chapter~5]{cfs}:
\bitem
\itemD The gauge group~$\U(1) \times \SU(2) \times \SU(3)$ of the standard model
with the correct couplings to the fermions. In more mathematical terms, the groups come with the correct representations acting on direct sums of spinorial wave functions.
\itemD Corresponding gauge fields and field equations. The~$\SU(2)$ gauge fields are left-handed and massive.
\itemD The Einstein equations, up to possible higher order corrections in curvature (which scale in powers of~$(\delta^2\: \text{Riem})$, where~$\delta \gtrsim \varepsilon$ is another regularization scale which can be identified with the Planck length and~$\text{Riem}$ is the
curvature tensor; for details see~\cite[Theorems~4.9.3 and~5.4.4]{cfs}).
\itemD In order to get deterministic equations in the continuum limit,
the number of generations must be equal to three.
\eitem

The first step in the continuum limit analysis is to write the kernel of the fermionic projector~\eqref{PC}
as a solution of a Dirac equation. This has the advantage that we can then perturb the system
simply by inserting bosonic potentials into the Dirac equation.
Clearly, the distribution in~\eqref{Psea} is a bi-solution of the Dirac equation. However,
a sum as in~\eqref{Pvac} no longer satisfies the Dirac equation. The way out is to rewrite
sums by direct sums (for details see~\cite[Section~2.3]{pfp} and~\cite[Section~3.4 and~\S4.2.6]{cfs}):
We first introduce the {\em{auxiliary fermionic projector}} by
\beq \label{Pauxdef}
P^\text{aux} = P^N_\text{aux} \oplus P^C_\text{aux}\:,
\eeq
where
\beq \label{Paux}
P^N_\text{aux} = \Big( \bigoplus_{\beta=1}^3 P^\text{vac}_{\tilde{m}_\beta} \Big) \oplus 0
\qquad \text{and} \qquad
P^C_\text{aux} = \bigoplus_{a=1}^7 \bigoplus_{\beta=1}^3 P^\text{vac}_{m_\beta} \:.
\eeq
Note that~$P^\text{aux}$ is composed of~$25$ direct summands, four in the neutrino
and~$21$ in the charged sector. The fourth direct summand of~$P^N_\text{aux}$ has the purpose
of describing the right-handed high-energy states; this gives more flexibility when introducing the
interaction below. Moreover, we introduce the
{\em{chiral asymmetry matrix}}~$X$ and the {\em{mass matrix}}~$Y$ by
\begin{align}
X &= \left( \1_{\C^3} \oplus \tau_\reg \,\chi_R \right) \oplus \bigoplus_{a=1}^7 \1_{\C^3} \label{asymmetry}\\
m Y &= \text{diag} \big( \tilde{m}_1, \tilde{m}_2, \tilde{m}_3, 0 \big)
\oplus \bigoplus_{a=1}^7 \text{diag} \big( m_1, m_2, m_3 \big) \:,
\end{align}
where~$m$ is an arbitrary mass parameter.
Here~$\tau_\reg \in (0,1]$ is a dimensionless parameter which has the purpose of keeping track
of the right-handed high-energy states. Introducing~$P^\text{aux}$ in this way has the advantage
that it satisfies the Dirac equation
\beq \label{dirvac}
(i \Pdd_x - m Y) \,P^\text{aux}(x,y) = 0
\eeq
(in order to keep the presentation simple, we assume that even the regularized objects
satisfy the Dirac equation; a more general treatment dropping this assumption is given
in~\cite[Chapter~5]{cfs}).
In order to introduce the interaction, we can now insert a general operator~$\B$ into the Dirac equation,\beq \label{Dinteract}
(i \Pdd_x + \B - m Y)  \,P^\text{aux}(x,y)  = 0 \:.
\eeq
The causal perturbation theory (see~\cite[Section~2.2]{pfp}, \cite{norm} or~\cite[Section~2.1]{cfs}
defines~$P^\text{aux}$ in terms of a unique perturbation series. The {\em{light-cone expansion}}
(see~\cite[Section~2.5]{pfp} and the references therein or~\cite[Section~2.2]{cfs}) is a method for analyzing
the singularities of~$P^\text{aux}$ near the light cone. This gives a representation of~$P^\text{aux}$
of the form
\beq \begin{split}  \label{lce}
P^\text{aux}(x,y) = & \sum_{n=-1}^\infty
\sum_{k} m^{p_k} 
{\text{(nested bounded line integrals)}} \times  T^{(n)}(x,y) \\
&+ \tilde{P}^\lec(x,y) + \tilde{P}^\hec(x,y) \:,
\end{split}
\eeq
where~$\tilde{P}^\lec(x,y)$ and~$\tilde{P}^\hec(x,y)$ are smooth to every order in perturbation theory.
Likewise, the nested bounded line integrals in~\eqref{lce} are smooth.
The factors~$T^{(n)}$, however, are singular on the light cone, and this singularity is regularized on the
scale~$\varepsilon$.
The auxiliary fermionic projector of the sea states~$P^\sea$ is obtained similar to~\eqref{Pauxdef} by
multiplication with the chiral asymmetry matrix.
Finally, we introduce the regularized kernel of the fermionic projector~$P(x,y)$ by forming the {\em{sectorial projection}} (see also~\cite[Section~2.3]{pfp} or~\cite[\S5.2.1]{cfs})
and by adding the contributions by the particle wave functions`$\psi_k$ and the anti-particle wave
functions~$\phi_l$,
\beq \label{partrace0}
(P)^a_b(x,y) = \sum_{\alpha, \beta} (\tilde{P}^\text{aux})^{(a,\alpha)}_{(b, \beta)}(x,y) 
-\frac{1}{2 \pi} \sum_k \psi^a_k(x) \overline{\psi^b_k(y)}
+\frac{1}{2 \pi} \sum_l \phi^a_l(x) \overline{\phi^b_l(y)}\:,
\eeq
where~$a,b \in \{1, \ldots, 8\}$ is the sector index, and the indices~$\alpha$ and~$\beta$ run over
the corresponding generations (i.e., $\alpha \in \{1, \ldots 4\}$ if~$a=1$ and~$\alpha \in \{1, 2, 3 \}$
if~$a=2, \ldots, 8$).

With these tools at our disposal, the continuum limit analysis can be carried out schematically as follows.
The operator~$\B$ in~\eqref{Dinteract} can be formed of general bosonic potentials
(electromagnetic, Yang-Mills, gravitational, chiral or even scalar or pseudo-scalar).
A-priori, these bosonic potentials can be chosen arbitrarily; in particular, they
do not need to satisfy any field equations.
Expressing the function~$\ell$ introduced in~\eqref{elldef} in terms of the regularized
kernel of the fermionic projector (using~\eqref{partrace0} and~\eqref{lce}),
the EL equations~\eqref{EL} become equations which involve the bosonic potentials
in~\eqref{Dinteract} (which enter the nested bounded line integrals in~\eqref{lce})
as well as the fermionic wave functions of the particles and anti-particles
(which enter in~\eqref{partrace0}).
Analyzing these equations asymptotically for small~$\varepsilon$, one finds that
the equations are satisfied if and only if the potentials in~$\B$ have a specific structure
and satisfy corresponding field equations (like for example Maxwell's equations).
These field equations are classical equations
for the bosonic potentials which involve the Dirac currents of the particles and anti-particles
as source terms. Moreover, the bosonic potentials couple to the wave functions
via the Dirac equation~\eqref{Dinteract}. In this way, one obtains a coupled interaction
described by classical bosonic fields.

\subsection{Connection to Quantum Field Theory} \label{secqft}
Going beyond the continuum limit, the interaction described by the causal action principle
can also be described by bosonic and fermionic quantum fields.
So far, these constructions have been carried out only for causal fermion systems
describing Minkowski space. More precisely, in~\cite{fockbosonic, fockfermionic}
it was shown that an interacting causal fermion system at a given time can be described
by a {\em{quantum state}}~$\omega$, a positive linear functional on the $*$-algebra of
observables~$\A$, i.e.\
\[ \omega^t \::\: \A \rightarrow \C \quad \text{complex linear} \qquad \text{and} \qquad
\omega^t \big( A^* A \big) \geq 0 \quad \text{for all~$A \in \A$}\:. \]
The observable algebra contains the usual physical observables like particle numbers,
densities, etc., and~$\omega^t(A)$ has the interpretation as the expectation value of this
observable at time~$t$ (for a pure state, this expectation value can be written in the usual
form~$\bra \Psi | A | \Psi \ket$ with a Fock vector~$|\Psi \ket$).
In~\cite{fockentangle} it was shown that this construction allows for the
description of general entangled states. The dynamics of this quantum state is
currently under investigation~\cite{fockdynamics}.

\subsection{Connection to Quantum Geometry and Quantum Gravity} \label{secqg}
The kernel of the fermionic projector~\eqref{Pxyabs} encodes geometric structures
giving rise to a mathematically concise setting of a Lorentzian quantum geometry~\cite{lqg}.
One important structure is the {\em{spin connection}}~$D_{x,y}$, being a unitary mapping (with respect to the spin inner product~\eqref{ssp}) between
the spin spaces,
\beq \label{Dxy}
D_{x,y} \::\: S_y \rightarrow S_x \qquad \text{unitary}.
\eeq
A first idea for construction~$D_{x,y}$ is to take a polar decomposition of~$P(x,y)$.
This idea needs to be refined in order to also obtain a metric connection and to ensure
that the different connections are compatible. Here for brevity we omit the details and refer
to~\cite{lqg} or the review~\cite{nrstg}. In general terms, it turns out that there is a canonical
spin connection~\eqref{Dxy}, provided that the operators~$x$ and~$y$ satisfy certain conditions,
which are subsumed in the notion that the spacetime points be {\em{spin-connectable}}.
{\em{Curvature}}~$\mathfrak{R}$ can be defined as the holonomy of the spin connection.
Thus, in the simplest case, for three points~$x,y,z \in M$ which are mutually spin-connectable,
one sets
\[ \mathfrak{R}(x,y,z) = D_{x,y}\: D_{y,z}\:D_{z,x} \::\: S_x \rightarrow S_x \:. \]
In~\cite{lqg} it is also shown that these structures give back the setting of Lorentzian spin geometry
as used in GR.

As outlined in Section~\ref{secdynamics}, the causal action principle describes the dynamics of our
quantum geometry. In this sense, causal fermion systems are a specific proposal for a theory of quantum gravity.
However, the connection to other approaches to quantum gravity (like canonical quantum gravity~\cite{kiefernew}
or loop quantum gravity~\cite{thiemann}) has not yet been worked out.

\section{Comparison} \label{sec:comparison} 
In this section we will highlight and compare the mathematical structures and the conceptional ideas behind
causal fermion systems and octonionic approaches to fundamental physics, and we will demonstrate a first simple example of how octonions can be integrated in causal fermion systems.
\subsection{Foundations of the Approaches}\label{sec:foundations}
We begin with a discussion of the basic ideas and the mathematical structures of the theories.

\subsubsection{Octonionic Approaches} 

Octonionic (division algebraic) approaches to the standard model remain an active research program, with different approaches considered in the literature. The underlying goal of these approaches is to find the correct algebraic structures, often in the form of tensor products of the division algebras, from which the gauge groups and particle content of the standard model may be derived, as well as the Lorentz group for the gravitational sector. An example of how the left associative multiplication algebra of the (complex) octonions may be used to represent one generation of standard model fermions with unbroken $SU(3)\times U(1)$ symmetry was given in \ref{Furey}. By further including the quaternion algebra $\bb{H}$ into this approach, it is possible to describe also the chiral weak interaction and the Lorentz spacetime symmetries \cite{dixon1990derivation,dixon2013division,furey2018demonstration,furey231}

Although such algebraic proposals provide an elegant derivation of the gauge symmetries and particle multiplets of the standard model, what is currently still lacking is a description of spacetime in which these particles live, as well as a dynamical framework. In the absence of a dynamical framework, these kinematic algebraic models cannot be rigorously tested against experimental data. For a review of an ongoing attempt to incorporate spacetime and dynamics see~\cite{singh-octo4}.

\subsubsection{Causal Fermion Systems}
Causal fermion systems are formulated on a set of operators~$\F^{\text{reg}}$ on a Hilbert space.
The measure~$\rho$ on~$\F^{\text{reg}}$
is the only ``dynamical'' degree of freedom in the theory. However, the fact that the support of the measure is typically restricted to a proper, lower dimensional subset of the operator manifold together with the intrinsic causal structures allows for the measure to encode a plethora of physical systems. In contrast to GR, for example, the operator manifold on which the variational principle is formulated does not fix the topology of the spacetime described by a minimizing measure. 

To describe a classical spacetime in terms of a causal fermion system we need to construct the fermionic projector~$P(x,y)$ which is a bi-distribution. In the coincidence limit this gives rise to the so-called local correlation map~$F: \mathcal{M}\rightarrow \F$ that allows us to identify points in classical spacetime with operators in~$\F$. The fermionic projector then allows us to calculate the causal action for a classical spacetime. An important observation for the following discussion is the fact that for every point~$x$, the spin space~$S_x$ is a~$2n$ dimensional vector space. Intuitively one can think of this vector space as a fiber space attached to a point~$x$ in~$\F$ that goes over to a fiber space of the classical spacetime in the continuum limit.

\subsubsection{Discussion}
In the following we will provide a first example of how octonions can be integrated in the mathematical structures of causal fermion systems and which conceptual lessons we can draw from this with respect to octonion based approaches and quantum gravity more generally. 

In view of~\eqref{Pvac} and~\eqref{PC} the vacuum is formed of eight direct summands.
In order to get a connection to the octonions, we replace the direct sum~\eqref{Pvac} by the octonion-valued ordinary sum
\beq
P(x,y) = P^N(x,y) \, e_0 + \sum_{i=1}^7 P^C_i(x,y)\: e_i \:. \label{PvacO}
\eeq
Here the index~$i$ at the factors~$P^C_i$ clarifies
that these factors could be different, describing the charged leptons
and the quarks of different colors and isospin. In the vacuum, the index~$i$ can be left out because,
as in~\eqref{PC}, all these factors are identical.
We note that the octonionic symmetry is broken, because~$e_0$ is distinguished
(as it is multiplied by~$P^N$, which is different from the factors~$P^C$).
We point out that the descriptions~\eqref{Pvac} and~\eqref{PvacO} are equivalent,
because the octonions simply serve as labels for the components of the direct sum in~\eqref{Pvac}.
Therefore, the crucial question is whether the formulation with octonions carries over and has
benefits in the description of interacting systems.

Before addressing this question in the next section, we point out that, with the above construction,
we implemented octonions into the causal fermion system {\em{of the vacuum}}.
Once this has been done, the interaction and the resulting dynamics are described by the
causal action principle. In this way, octonions should be helpful for getting a better understanding of
and giving an explanation for the choice of the vacuum configuration in causal fermion systems.

In interacting causal fermion systems obtained from an octonionic vacuum like~\eqref{PvacO}
(using the methods outlined in Sections~\ref{seccfssm} and~\ref{secqft}), the octonionic structures are still 
present in the {\em{local}} form of the causal fermion system.
More precisely, the octonionic structures describe the internal degrees of freedom contained in
the fibers of vector bundles (like the spin space~$S_x$, linear operators thereon 
like the kernel of the fermionic projector~$P(x,y) : S_y \rightarrow S_x$, etc.).
Since octonionic theories by themselves do not provide spacetime structures and dynamical equations,
this is the best we can hope for without imposing additional structures.
It is interesting to note that in~\cite{singh-octo2, singh-td} it was proposed to use Adler's trace dynamics~\cite{adler-trace} and Connes' spectral action principle~\cite{connes}
in the octonionic setting. In an upcoming paper~\cite{tracedynamics} we will show that the causal action principle reduces to trace dynamics on the diagonal (i.e.\ if only local contributions~$\L(x,x)$ are considered). This further strengthens the above claim that octonion based algebraic structures are suitable to describe physics locally,
i.e.\ in fibers of corresponding vector bundles.

\subsection{Connection to the Gauge Group of the Standard Model}\label{sec:gaugegroup}
Obtaining the gauge group of the standard model is a key objective for any fundamental physical theory. Here we summarize the current status of this quest in both approaches and discuss
what we can learn from a comparison. 

\subsubsection{Octonionic Approaches} 
Algebraic approaches based on the octonions (or more generally the division algebras) attempt to recover the standard model gauge groups as those groups that preserve certain subspaces of the underlying algebraic structure. These subspaces are identified by means of projectors, as discussed in \ref{historical}.

Different constructions are possible. Dixon takes as the spinor space the algebra ~$\bb{T}=\mathbb{R}\otimes\mathbb{C}\otimes\mathbb{H}\otimes\mathbb{O}$ \cite{dixon2013division}. Such spinors are acted upon from the left or right by the algebra itself, with these linear maps generating the associative multiplication algebra. In this construction it is the nontrivial algebraic structure of the spinor space itself that provides the algebraic source of the internal symmetries needed to replicate the structure of the standard model. The projectors that may be identified within  ~$\bb{T}$ simultaneously decompose the spinor space into mutually orthogonal subspaces, as well as the associative multiplication algebra into subalgebras whose actions preserve these subspaces. 

Furey's approach is similar, but identifies the algebraic spinors are the minimal left ideals of the left associative multiplication algebra of $\bb{C}\otimes\bb{O}$ (see \ref{Furey}) and $\bb{C}\otimes\bb{H}\otimes\bb{O}$ \cite{furey2016standard,furey231}. The dimensionality of these minimal left ideals then tells one the number of distinct physical states, whereas the gauge symmetries are those symmetries that preserve the minimal ideals.

\subsubsection{Causal Fermion Systems}
Assuming the fermionic projector~$P(x,y)$ in vacuum to be given by expression~\eqref{Pvac} with eight direct summands, one can prove that the Minkowski vacuum is a minimizer of the causal action principle in the limit where~$\varepsilon \searrow 0$. In this case~$P(x,y)$ satisfies the Dirac equation~\eqref{dirvac}. If we perturb the fermionic projector to satisfy the Dirac equation with a bosonic potential~\eqref{Dinteract} we find that this perturbation only corresponds to a minimizer of the causal action if the vector potential~$B$ has the right transformation property under the gauge group of the standard model.

\subsubsection{Discussion}
We start the discussion in this section with the question how the octonionic vacuum~\eqref{PvacO} can be perturbed.
The auxiliary fermionic projector can be defined again by replacing direct sums by sums
and inserting the octonions as factors, i.e.
\[ P^\text{aux} := P^N_\text{aux} \: e_0 + \bigg( \bigoplus_{\beta=1}^3 P^\text{vac}_{m_\beta} \bigg)
\big( e_1 + \cdots + e_7 \big) \:. \]
Now the vacuum Dirac equation~\eqref{dirvac} holds if the matrix~$Y$ is represented by
a linear operator acting on the octonions. The interaction can again be
described by inserting an operator~$\B$ into the Dirac equation~\eqref{Dinteract}.
Now we observe that for the equation~\eqref{Dinteract} to be well defined,
the operator~$\B(x)$ involves operations on the octonions.
This leads us to choose the matrix elements of~$\B$ 
as being composed of elements of the algebra of endomorphisms acting on the octonions.
This algebra is the left associative multiplication algebra~$\bb{O}_L\cong \bb{C}\ell(6)$. This
algebra is associative. Moreover, this algebra acts on
the octonions by left multiplication, giving a mapping
\[  \bb{O}_L \times \mathbb{O} \rightarrow \mathbb{O} \:. \]
In summary, when integrating octonions in the construction of the fermionic projector one is naturally lead to look for the standard model in the left associative multiplicative algebra. This is one of the key insights in this paper as it provides a clear motivation as for why one would consider this algebra instead of the octonions themselves.
Furthermore, integrating the octonions in the fermionic projector provides a clear road map of how to get to the standard model. On the other hand, working with octonions provides a clear motivation for the choice of the vacuum in causal fermion systems which otherwise might appear arbitrary. 

\subsection{Describing the Three Generations in the Standard Model} \label{secthreegen}
We now explain and compare how the three generations of elementary particles
(like~$e$, $\mu$ and~$\tau$) come up in the different approaches.
\subsubsection{Octonionic Approaches}
Although existing division algebraic approaches provide an elegant construction of the internal space of a single generation of leptons and quarks, a comprehensive algebraic explanation for the existence of three generations remains to be found\footnote{It is worth noting that most GUTs likewise inherently correspond to single generation models, lacking any theoretical basis for three generations, which ultimately has to be imposed by hand.}.

In~\cite{manogue1999dimensional} three generations of leptons are described in terms of the three~$\bb{H}$ subalgebras of~$\bb{O}$ that contain a particular imaginary octonion unit. 

Furey identifies three generations of color states directly from the 64 complex dimensional algebra~$\bb{C}\ell(6)$~\cite{furey2014generations}. Constructing two representations of the Lie algebra~$\mathfrak{su}(3)$, the remaining 48 degrees of freedom transform under the action of the~$\SU(3)$ as three generations of leptons and quarks. However, the inclusion of~$\U(1)_{em}$ via the number operator, which works in the context of a one-generation model, fails to assign the correct electric charges to states in this three-generation model. A generalized action that leads to a generator that produces the correct electric charges for all states is introduced in~\cite{furey2018three}.

Dixon instead describes three generations in terms of the algebra~$\bb{T}^6=\bb{C}\otimes\bb{H}^2\otimes\bb{O}^3$, with a single generation being described by~$\bb{T}^2$, a complexified (hyper) spinor in 1+9D spacetime~\cite{dixon2004division}. The choice~$\bb{T}^6$, as opposed to any other~$\bb{T}^{2n}$ appears rather arbitrary however, although can be motivated from the Leech lattice.

One intriguing idea is that the triality automorphism of~$\Spin(8)$, may be responsible for the existence of three generations. Triality corresponds to an outer automorphism of~$\Spin(8)$ that permutes between the vectors and two spinor representations, all of which are of dimension eight. It is known that there exist three conjugacy classes of~$\Spin(7)$ subgroups in~$\Spin(8)$ which are permuted by triality~\cite{varadarajan2001spin}. Furthermore, there exists a unique way to choose one~$\Spin(7)$ subgroup from each conjugacy class such that the common intersection of these subgroups is~$G_2$. This suggests it may be possible to represent three generations of color states (since~$G_2$ contains~$\SU(3)$ as one of its maximal subgroups) in terms of~$\Spin(8)$, with each generation corresponding to a~$\Spin(7)$ subgroup. $\Spin(8)$ corresponds to the spin group of~$\bb{C}\ell(8)$, whose bi-vectors generate~$\mathfrak{spin}(8)$. One way to generate~$\bb{C}\ell(8)$ is to consider the left or right associative multiplication algebra of a column vector of two octonions~\cite{schray1994octonions}. 

An alternative way to generate~$\bb{C}\ell(8)$ is to instead consider the left or right associative multiplication algebra generated from sedenions~$\bb{S}$, which may be generated from the octonions\footnote{Starting with~$\mathbb{R}$, the remaining three division algebras can be generated via what is called the Cayley-Dickson process. This process does not terminate with~$\bb{O}$ however, but continues indefinitely to produce a series of $2^n$-dimensional algebras.}. It has recently been advocated that~$\bb{S}$, constitutes a natural mathematical object which exhibits the required algebraic structure necessary to describe three generations of color states~\cite{gillard2019three,gresnigt2019sedenions,gresnigt2023three}. In this construction, one generation of leptons and quarks is represented in terms of two minimal left ideals of~$\bb{C}\ell(6)\subset \bb{C}\ell(8)$, in what corresponds to a direct generalization of the octonionic construction in~\cite{furey2016standard}, reviewed in Subsection~\ref{Furey}. Subsequently the~$S_3$ automorphism of order three, which is an automorphism of~$\bb{S}$ but not of~$\bb{O}$ can be used to generate two additional pairs of minimal~$\bb{C}\ell(6)$ ideals to represent exactly two additional generations of color states. Exactly how the~$S_3$ automorphisms of~$\bb{S}$ are related to the~$S_3$ outer automorphisms corresponding to triality remains to be worked out, as does a possible link between this sedenionic construction and the three conjugacy classes of~$\Spin(7)$ subgroups of~$\Spin(8)$ mentioned above.  

Finally, numerous authors have encoded three generations within the exceptional Jordan algebra~$J_3(\bb{O})$ consisting of three by three matrices over~$\bb{O}$~\cite{dubois2016exceptional,dubois2019exceptional,todorov2018octonions,todorov2018deducing,boyle2020standard2,boyle2020standard}. In these models, the three octonions in~$J_3(\bb{O})$ are related via triality. Each octonion is associated with one generation via the three canonical~$J_2(\bb{O})$ subalgebras of~$J_3(\bb{O})$. The automorphism group of~$J_3(\bb{O})$ is the exceptional group~$F_4$. The standard model gauge group emerges as the intersection of two subgroups of~$F_4$, the first being the subgroup that preserves the representation of the octonions occurring in the elements of~$J_3(\bb{O})$ as elements of~$\bb{C}\oplus\bb{C}^3$, and the second being~$\Spin(9)$, the automorphism group of each~$J_2(\bb{O})$ subgroup of~$J_3(\bb{O})$.

\subsubsection{Causal Fermion Systems}
In the ansatz~\eqref{Pvac}, \eqref{PC} and~\eqref{massneutrino} the three generations of
elementary particles are built in ad hoc via the $\beta$-sum. At this stage, one can just as well
choose a different number of generations~$g \in \N$ by summing instead over~$\beta=1,\ldots, g$.
However, the analysis of the continuum limit reveals that one gets field equations with a well-posed
Cauchy problem if the number of generations is three. In simple terms, in the case~$g<3$ the resulting
equations are overdetermined (meaning that they do not admit non-trivial solutions), whereas in the
case~$g>3$ they are under-determined (meaning that, given initial data the solution is not unique).
The details of this analysis and a discussion of there results can be found in~\cite[Section~3.7]{cfs}.

\subsubsection{Discussion}
In order to make a connection between causal fermion systems and octonionic approaches,
it is preferable to write out the direct sum over the generation index in~\eqref{Paux} as a matrix,
\[ \bigoplus_{\beta=1}^3 
{P^\text{sea}_{m_\beta}} = \begin{pmatrix} P^\text{sea}_{m_1} & 0 & 0 \\
0 & P^\text{sea}_{m_1} & 0 \\ 0 & 0 & P^\text{sea}_{m_3} \end{pmatrix} \:, \]
giving a $3 \times 3$-matrix with octonionic entries.
The form of the interaction poses constraints on the form of these matrices.
The interesting connection with octonionic theories is that specific algebras arising in this context,
like the exceptional Jordan algebra~$J_3(\bb{O})$ in the approach~\cite{singh-jordan, singh-octo3}, give concrete proposals for how these constraints could look like. 
The hope is that studying the causal action principle for
these proposals will give a better understanding of why the gauge groups of the standard model appear in nature
and why the gauge fields couple to matter in the way observed in experiments.
This will be discussed in more detail in Section~\ref{sece68}.

\section{Outlook and Directions of Future Research} \label{secoutlook}
In this section, we given an outlook and explain directions of future research.

\subsection{Connection with the Exceptional Lie Groups and Jordan Algebras} \label{sece68}
In Section~\ref{secthreegen} we saw that describing the interactions of the standard model
in the setting of causal fermion systems naturally leads to the exceptional Jordan algebra.
Moreover, it brings with it the exceptional Lie groups which have a direct connection with the octonions.
We now outline these connections, also pointing towards directions of future research.
There are five exceptional groups, labeled~$G_2, F_4, E_6, E_7$ and~$E_8$. The smallest of these five groups is the $14$-dimensional~$G_2$ which is the automorphism group of the octonions. The $52$-dimensional Lie group~$F_4$ is the automorphism group of the exceptional Jordan algebra which is the algebra of $(3\times 3)$ Hermitian matrices with octonionic entries. The next larger group~$E_6$ is $78$-dimensional, and is the automorphism group of the complexified Jordan algebra. The group~$E_7$ is $133$-dimensional, and the largest exceptional group~$E_8$ is $248$-dimensional. The subgroups of these exceptional groups have striking similarities to the gauge groups
of the standard model. While it is true that these are rather large groups and exhibit very many subgroups and branchings apart from the standard model, that fact alone does not take away their potential significance. If we demand that the algebraic structure of the associated Lie algebras should also explain the values of the fundamental constants of the standard model, that challenge is enough to rule out most branchings and we are constrained to find at least one branching which successfully explains the experimentally observed data of the standard model.
 As one promising example consider starting from the smallest exceptional group~$G_2$, which has a maximal subgroup~$\SU(3)$. The generators of the Clifford algebra~$\bb{C}\ell(6)$ constructed from octonionic chains exhibit an $\SU(3)$-symmetry which can be identified with~$\SU(3)_{C}$ of QCD because it correctly describes one generation of standard model quarks and leptons. This could be any one of the three known fermion generations, but~$G_2$ is not large enough to accommodate all three generations. Consider however the next exceptional group~$F_4$ which admits a subgroup~$\SU(3)\times \SU(3)$~\cite{Yokota}. We can associate the first~$\SU(3)$ with three fermion generations and the second~$\SU(3)$ with~$\SU(3)_{C}$ as already done with~$G_2$. This is consistent with the already mentioned observation that~$F_4$ is also the automorphism group of~$3\times 3$ Hermitian matrices with octonionic entries.  These matrices describe three generations of standard model fermions and their characteristic equation (a cubic with three distinct real roots) shows evidence for determining the fundamental constants of the standard model~\cite{singh-octo3}.
 
 The next larger group~$E_6$ admits the subgroup~$\SU(3)\times \SU(3) \times \SU(3)$ and now, the newly introduced~$\SU(3)$ admits the branching~$\SU(2) \times \U(1)$ which is identifiable with the electroweak sector. Finally, the group~$E_8$ admits the subgroup~$\SU(3) \times E_6$, and with~$E_6$ already able to account for the standard model in the above mentioned branching, the newly introduced~$\SU(3)$ here can be identified with an octonionic vector space on which the standard model particle states are defined, and from which our 4D spacetime is assumed to be emergent~\cite{Kaushik, singh-octo4}.
 
These attractive properties of the exceptional Lie groups in connection to the octonions encourage further investigation of their role in the unification of interactions.  

\subsection{Chiral Symmetry Breaking} \label{secchiralsymm}
As explained in Section~\ref{seccfssm}, in the causal fermion system
description of the standard model, the chiral symmetry is broken by hand via the ansatz~\eqref{asymmetry}.
This procedure is not quite satisfying. Instead, within the context of causal fermion systems, it would be desirable to explain the chiral symmetry breaking
from the interactions as described by the causal action principle. This should be achieved in a manner consistent with octonionic theories where it has recently been proposed that algebraic symmetry breaking on a physical system described by a tensor product of division algebras can be induced via the identification of certain complex structures within the (tensor product of) division algebras \cite{furey231}.

With regards to the causal action principle, we could generalize~\eqref{asymmetry} or replace it by an alternative ansatz which does
not break the chiral symmetry ad hoc. Using the methods introduced in~\cite{cfs}, one can analyze
the dynamics of the resulting systems in the continuum limit.
This procedure also opens the pathway to phenomenological considerations that extend to both the standard model gauge interaction sectors and the gravitational sector. We envisage three main scenarios, depending on whether the chiral symmetry breaking happens in the gauge sectors related to the standard model, or in the gravity sector, or in both these sectors.

\begin{enumerate}[leftmargin=2em]
\item
It is natural to connect the breakdown of the exact left-right symmetry to a spontaneous violation of parity that is either instantiated as an asymmetric minimization of the potential, or can be traced back to a dynamical origin implemented through the loop corrections to the potential. Senjanovic and Mohapatra considered in~\cite{Senjanovic:1975rk}, as an example of this mechanism, a theory with gauge group \SU(2)$\!\!\!\phantom{a}_L\times$\SU(2)$\!\!\!\phantom{a}_R\times$U(1), unifying weak and electromagnetic interactions, with extended Higgs sector defined by the multiplets~$\Phi_L$ and~$\Phi_R$, respectively in the representations~$(\frac{1}{2},0,1)$ and~$(\frac{1}{2},0,1)$. The multiplets fulfill the relation~$Q=T^3_L+T^3_R+\frac{1}{2}Y$, with~$Q$ electric charge, $Y$ hypercharge and~$T^3_{L ,R}$ third component of the isospin for the \SU(2)$\!\!\!\phantom{a}_L$, \SU(2)$\!\!\!\phantom{a}_R$ sector. This implies that
\[ \Phi_{L ,  R}= \begin{pmatrix} \Phi^{\dagger}_{L ,  R} \\ \Phi^0_{L ,  R} \end{pmatrix} \:, \]
where the left-right discrete symmetry is manifest. Taking then into account a general potential for~$\Phi_L$ and~$\Phi_R$ that does not spoil renormalizability and gauge invariance and is left-right symmetric, i.e.  
\begin{align*}
V(\Phi_L, \Phi_R) &= -\mu^2 \left( \Phi_L^\dagger \Phi_L + \Phi_R^\dagger \Phi_R \right) + \lambda_1 \left[ (\Phi_L^\dagger \Phi_L)^2 + (\Phi_R^\dagger \Phi_R)^2 \right] \nonumber \\
&\quad\: +\lambda_2 (\Phi_L^\dagger \Phi_L) (\Phi_R^\dagger \Phi_R)\:,
\end{align*}
it was demonstrated, inspecting the equations for the minima, i.e.\ $\partial V/\partial \Phi^\dagger_L=0$ and~$\partial V/\partial \Phi^\dagger_R=0$, the existence of asymmetric solutions. These correspond to a range of values of the real parameters~$\mu^2$, $\lambda_{1,2}$ for which~$\langle \Phi_L\rangle=0$ and~$\langle \Phi_R\rangle=u_R\neq 0$, the value of~$u_R$ being determined by the constraint~$-\mu^2+2\lambda_1 u_R^2=0$, yielding
\[ u_R=\left( \frac{\mu^2}{2 \lambda_1} \right)^{\frac{1}{2}} \:. \]
The solutions with lowest energy that violate parity then corresponds to the parameter space~$\mu^2>0$ and~$\lambda_2 > 2 \lambda_1$, providing a model with spontaneous parity violation. This solution can be shown~\cite{Senjanovic:1975rk} to arise as a higher order effect, following the perspective of Coleman and Weinberg~\cite{Coleman:1973jx}. This can be immediately recognized if one simply focuses on the \U(1)$\!\!\!\phantom{a}_L\times$\SU(1)$\!\!\!\phantom{a}_R$ gauge group, taking into account only the two multiplets~$\phi_L$ and~$\phi_R$, respectively in the representations~$(1,0)$ and~$(0,1)$. The effective potential in the one loop approximation was therefore provided in~\cite{Senjanovic:1975rk}, extending the results of~\cite{Coleman:1973jx}, by the expression 
\begin{align*}
\qquad \quad V(\phi_L, \phi_R) &= \lambda_1 \left[ (\phi_L^\star \phi_L)^2 + (\phi_R^\star \phi_R)^2 \right] + \lambda_2 (\phi_L^\star \phi_L)(\phi_R^\star \phi_R) \\
&\quad\:+ \frac{3 g^4}{64 \pi^2} (\phi_L^\star \phi_L)^2 \!\! \left[ \ln \frac{\phi_L^\star \phi_L}{M^2} -\frac{25}{6}\right] 
\!\!+\!\! \frac{3 g^4}{64 \pi^2} (\phi_R^\star \phi_R)^2 \left[ \ln \frac{\phi_R^\star \phi_R}{M^2} -\frac{25}{6}\right] \\
&\quad\: + \O(\lambda_1^2, \lambda_1g^2, \dots)\:,
\end{align*}
recovered in the Landau gauge and with~$\lambda_1\sim g^2$. For normalized couplings within the range~$\lambda_2 > 3 g^2 / 64 \pi^2$ the potential has the minima~$\langle\phi_L \rangle=0$ and~$\langle\phi_R \rangle^2=v_R\neq 0$, where~$v_R$ fulfills the constraint
\[ \ln \frac{v_R^2}{M^2}   -\frac{25}{6} = -\frac{1}{2} - \frac{64 \pi^2}{3 g^4} \lambda_1 \:. \]

It is natural to extend the same mechanism considered in~\cite{Senjanovic:1975rk} to any sector of the standard model, taking into account the mirror images of the chiral-symmetry broken content of matter that is present in the Universe, as first suggested by Lee and Yang in~\cite{Lee:1956qn}. Mirror models~\cite{MDM1,MDM2,MDM3, MDM4,MDM5,MDM6,MDM7,MDM8} have then been proven to provide a rich phenomenology~\cite{MDMP1,MDMP2,MDMP3}, including the possibility to account for dark matter~\cite{MDMPDM1,MDMPDM2}. A minimal realization of this scenario certainly encodes the dark photon~\cite{dpA,dpB,dpC}, a model yielding rich phenomenological consequences in the multi-messenger perspective~\cite{dp18}. The model encodes the Lagrangian for the hidden sector  
\[ \mathcal{L}=\mathcal{D}_\mu s^\dagger \mathcal{D}^\mu s + \bar{\chi} (\imath \gamma^\mu \mathcal{D}_\mu - \mu_\chi ) \chi -\frac{1}{4} F'_{\mu \nu} {F'}^{\mu \nu}  -\frac{\varepsilon}{2} F_{\mu \nu} {F}^{\mu \nu} + V(s,\chi) \:, \]
where~$F'_{\mu\nu}$ denotes the field-strength of the dark photon~${A'}_\mu$, $F_{\mu\nu}$ denotes the field-strength of the photon~$A_\mu$, $\varepsilon$ is a technically naturally small parameter that introduces a mixing term among the two gauge sectors, $\mathcal{D}_\mu =\partial_\mu + \imath g' A'_\mu$ is the covariant derivative 
\Felix{Should this better be $\mathcal{D}_\mu =\partial_\mu + \imath g' A'_\mu + \imath g A_\mu$?}%
with respect to the gauge photon, with~$g'$ coupling constant in the dark sector, $s$ is a scalar singlet and~$\chi$ 
denotes a Dirac fermion charged under the dark~$\U'(1)$, but transforming like a singlet under the standard model gauge sector. An effective non-renormalizable potential related to a not-unifying theory can be claimed to be responsible for the spontaneous symmetry breaking of the dark $\U'(1)$-symmetry, e.g. 
\[ V(s,\chi)= y' s \bar{\chi}\chi + m_s^2 s^\dagger s + \frac{\lambda_s}{4} (s^\dagger s)^2 + \frac{1}{\Lambda^2} (s^\dagger s)^3 +\cdots \]
with~$y'$ Yukawa coupling between the~$s$ and~$\chi$ fields. Selecting the vacuum state~$\langle s \rangle=v_s$, with~$v_s^2=- 4 m_s^2/\lambda_s$, induces the spontaneous breakdown of the $\U'(1)$-symmetry, and then provide the mass~$m_{A'}^2={g'}^2 v_s^2$ to the dark photon. The dimension six operators are responsible for first order phase transitions, which in turn leave an imprinting in the stochastic background of gravitational waves of cosmological origin~\cite{dp18}. A dynamical mechanism of the type investigated in~\cite{Senjanovic:1975rk} could be investigated as being responsible for the spontaneous symmetry breaking that provides a mass to the dark photon.

Another phenomenologically relevant study-case is offered by the Majoron mo\-del. This introduces a~$\U(1)$ global symmetry that account for the conservation of the baryon-lepton number. The symmetry is then spontaneously broken by a complex scalar field that is coupled to neutrinos and to the Higgs particle. This mechanism then entails additional terms to the total Lagrangian that are beyond the standard model, namely  
\[ \mathcal{L}_{\rm Majoron }= f H \bar{L} \nu_R + h \sigma \bar{\nu}_R \nu_R^c + h.c. + V(\sigma, H) \:, \]
where~$\sigma$ denotes the complex scalar field 
\[ \sigma=\frac{1}{\sqrt{2}}(v_{\rm B-L}+ \rho +\imath \varphi) \:, \]
$h$ and~$f$ are Yukawa matrices, and 
\begin{equation}\label{Vs}
V(\sigma, H)=V_0(\sigma, H)+V_1(\sigma)+V_2(h,\sigma)\,.
\end{equation}
In~\eqref{Vs}, the renormalizable dimension four operators read
\begin{align*}
V_0(\sigma, H) &= \lambda_{\sigma} \left( |\sigma|^2 -\frac{v_{\rm B-L}^2}{2} \right)^2  +  \lambda_H \left( |H|^2 -\frac{v^2}{2} \right)^2 \\ 
&\quad\:+ \lambda_{\sigma H}  \left( |\sigma|^2 -\frac{v_{B-L}^2}{2} \right) \left( |H|^2 -\frac{v^2}{2} \right) ,
\end{align*}
the dimension five self-interaction operators for the complex scalar~$\sigma$, suppressed by the new physics energy scale~$\Lambda$, are expressed by
\[ V_1(\sigma)= \frac{\lambda_1}{\Lambda} \sigma^5 +  \frac{\lambda_2}{\Lambda} \sigma^\star \sigma^4 +  \frac{\lambda_3}{\Lambda} (\sigma^\star)^2 \sigma^3 + {\rm h.c.}\,, \]
and finally dimension six operators that eventually realize first order phase transitions~\cite{Addazi:2017oge} acquire the form
\[ V_2(H,\sigma)= \beta_1 \frac{(H^\dagger H)^2\sigma}{\Lambda} +  \beta_2 \frac{(H^\dagger H)\sigma^2 \sigma^\star}{\Lambda} +  \beta_3 \frac{(H^\dagger H)\sigma^3 }{\Lambda} + {\rm h.c.}\:. \]
The Majoron corresponds to the pseudo-scalar field~$\varphi$, a Nambu-Goldstone boson that emerges due to the spontaneous breakdown of the~$\U(1)$ symmetry. 

The model can be naturally connected to neutrino mass-generation~\cite{Addazi:2019dqt}, while accounting for a multiplicity of phenomenological instantiations. Due to the spontaneous symmetry breaking of the global~$\U(1)$ symmetry, RH neutrinos acquire a Majorana mass 
\[ M= \frac{1}{\sqrt{2}} h v_{\rm B-L} \:, \]
while LH neutrinos acquire a Dirac mass
\[ m= \frac{1}{\sqrt{2}} f v \:. \]
The model naturally implements a see-saw mechanism based upon the hierarchy~$M\!>\!\!>\!m$; namely,
\[ N= \nu_R + \nu_R^c +\frac{m}{M} (\nu_L + \nu_L^c)\,,
\nu= \nu_L + \nu_L^c -\frac{m}{M} (\nu_R + \nu_R^c)\,, \]
which finally provide~$m_N\simeq M$ and~$m_\nu\simeq m^2/M$, for~$m_N$ mass of the right-handed neutrino~$N$. 

Gauging the dark photon with the Majoron also provides an interesting possibility for phenomenology. This is in principle a scenario that enables to encode the paradigm developed in~\cite{Senjanovic:1975rk}, but the exact implementation of which will anyway require a detailed study.

\item
Parity being violated at cosmological scales is a fascinating possibility that, when proven real, would be ground-breaking as much as it was the discovery by Chien-Shiung Wu of parity violation in the weak interactions, later theoretically accounted for in the model developed by Lee and Yang in~\cite{Lee:1956qn}. Theoretically, the breakdown of the chiral symmetry induces in the gravitational field an imbalance among the amplitudes of the chiral components of the phase-space variables. A prototype of this unbalance is due to the inclusion in the action of gravity of a Pontryagin term, which plays the role of the $\theta$-sector in Yang-Mills theory, notably in QCD. Within this latter, the effect cannot be classical, as the Pontryagin term, topological, does not affect the equations of motion. Conversely, the $\theta$-term affects the vacuum of the quantum theory, modifying its topology. For the gravitational field, instead, parity violation at the classical level can only be introduced dynamically, coupling the gravitational Pontryagin term with a dynamical (pseudo-)scalar field or a fermionic bilinear, i.e.
\Felix{What is the $R_{\mu \nu}$ here? It's a field strength, right?}
\begin{equation}  \label{lagrav}
\mathcal{S}_{\rm gravity} = \mathcal{S}_{\rm EH} + \frac{\alpha}{4} \int \phi R_{\mu \nu} R_{\rho \sigma} \epsilon^{\mu \nu \rho \sigma}\,,
\end{equation}
having denoted with~$\mathcal{S}_{\rm EH}$ the Einstein-Hilbert action, with~$\epsilon$ the Levi-Civita symbol, with~$\phi$ either a scalar or a pseudo-scalar field, and with~$\alpha$ a scale with inverse dimension of mass. The inclusion of a pseudo-scalar field does not spoil the parity symmetry of the theory. Nonetheless, when the pseudo-scalar field evolves, for instance dynamically, towards its vacuum state, the chiral symmetry of gravity is manifestly broken. Chiral symmetry is dynamically broken for a background (homogeneous) solution of~$\phi$ on a Friedmann-Lema\^itre-Robertson-Walker (FLRW) metric of the type~$ds^2=dt^2-a^2(t)d\vec{x}^2$. Here, the non-vanishing components of the gravitational perturbations can be expressed in terms of the of the left/right-circular polarization basis 
\begin{eqnarray}
&h_{11}= - \frac{1}{\sqrt{2}} (h_L + h_R)\,, \qquad h_{22}=  \frac{1}{\sqrt{2}} (h_L + h_R) \nonumber\\
&h_{12}=h_{21}=  \frac{i}{\sqrt{2}} (h_L - h_R) \nonumber\,,
\end{eqnarray}
fulfill the equations
\begin{eqnarray}
& \bar{\Box} h_R= i \frac{16 \pi G}{a} \left( \ddot{\bar{\phi}} -  H\, \dot{\bar{\phi}}  \right)  \partial_z \dot{h}_R \
 \nonumber\\
& \bar{\Box} h_L= - i \frac{16 \pi G}{a} \left( \ddot{\bar{\phi}} -  H\, \dot{\bar{\phi}}  \right)  \partial_z \dot{h}_L \
 \nonumber\,,
\end{eqnarray}
having denoted with~$\bar{\Box}$ the D'Alembertian operator on the FLRW metric, with~$\bar{\phi}$ the dynamical solution for~$\phi$ consistent with the FLRW background, with dots derivatives with respect to the time~$t$, and introduced the Hubble parameter~$H=\dot{a}/a$. The dynamics of the scalar field can be thought to be generated at the quantum level --- the path integral for he scalar field will entail a kinematic term for it. Alternatively, a free Lagrangian for~$\phi$ can be added to~\eqref{lagrav}, i.e.
\[ 
\mathcal{S}_{\rm free}^{\phi} = \int d^4 x \sqrt{-g} \left( g^{\mu \nu} \partial_\mu \phi \partial_\nu \phi + m^2_\phi \, \phi^2 \right)\:. \]
The Pontryagin term (improperly called in the literature ``gravitational Chern-Simons'' term) can be recovered from the chiral anomaly, evaluated on a graviton condensate~\cite{Schwartz, Fujikawa} --- the gravitons condensate arises from considering a fixed background of the gravitational field strength, or in other words constant eclectic and magnetic gravitational components.   
For this purpose, we may combine~$\partial_\mu J_5^\mu= 2 i m \,  \bar{\psi} \gamma^5 \psi$, which holds due to the equation of motion for~$J_5^\mu=\bar{\psi}  \gamma^5 \gamma^\mu \psi$, with the chiral anomaly calculated in a coherent state of gravitons~$| h\rangle$, i.e. 
\[ \langle h | \partial_\mu J_5^\mu | h \rangle = - \frac{1}{384 \pi^2}  R_{\mu \nu} R_{\rho \sigma} \epsilon^{\mu \nu \rho \sigma}\ \:, \]
hence obtaining 
\[ \langle h |   \bar{\psi} \gamma^5 \psi | h \rangle \simeq  R_{\mu \nu} R_{\rho \sigma} \epsilon^{\mu \nu \rho \sigma}\:. \]
The gravitational Chern-Simons term in~\eqref{lagrav}, may then originate from a Yukawa coupling term of the form
\[ \mathcal{L}_{\rm Yukawa} = y \, \phi\,  \bar{\psi} \gamma^5 \psi \:. \]
A dynamical solution for~$\phi$ would then dynamically break the chiral symmetry of the gravitational sector, with possible observational consequences~\cite{alexander2022chernsimons,crequesarbinowski2023parityviolating}. 

Let us consider now a step forward in this direction, assuming that the very same gauge fields emerge as fermionic condensates. We may have two possibilities: i) the fermionic fields that undergo condensation are the standard model ones, and no extra matter degrees of freedom are requested; ii) extra fermionic fields are requested, as a technicolor sector of the gauge interactions. The second option opens the pathway to a plethora of possibilities, which we would entail eventually a loss of falsifiability of the theory. Let us suppose then to work within the first working assumption.

To address condensation, we consider the mechanism proposed by Bjorken~\cite{Bjo1, Bjo2, Bjo3} to obtain gauge fields from the Nambu--Jona-Lasinio condensation mechanism~\cite{NJL1, NJL2}. We may start from the Nambu--Jona-Lasinio Lagrangian density
\[ 
\mathcal{L}_{\rm NJL}=\bar{\psi} \left( i \Pdd -m \right) \psi + \frac{G}{2} \left( \bar{\psi} \gamma_\mu \psi  \right)^2, \]
with~$G$ coupling constant of the self-interaction, and then write the generating functional path integral
\Felix{What is $\psi^\mu$? Should it be~$\partial_\mu \psi$?}%
\begin{align*} 
\mathcal{W}_{\rm NJL}[J] &=\int \mathcal{D}\bar{\psi}\,  \mathcal{D}\psi e^{ i \int d^4 x \left[  \mathcal{L}_{\rm NJL} - \bar{\psi} \gamma_\mu \psi ^\mu  \right]} \\ 
&= N \int \mathcal{D}\bar{\psi}\,  \mathcal{D}\psi \, \mathcal{D}A \, e^{ i \int d^4 x \left[  \mathcal{L}_{\rm NJL} - \bar{\psi} \gamma_\mu \psi ^\mu  -\frac{1}{2G} (A_\mu -G \bar{\psi}\gamma_\mu \psi )^2\right]} \,,
\end{align*}
where in the last step the integration over an auxiliary Gaussian factor was introduced, yielding multiplication by a constant. One can immediately notice that~$-iG g_{\mu \nu}$ plays the role of the bare photon propagator~$i e_0^2 g_{\mu \nu}/q^2$.

At the Lagrangian density level, the auxiliary field~$A$ is still a Lagrangian multiplier. Nonetheless, at the path integral level, integration over the fermionic fields will turn the auxiliary field~$A$ into a dynamical one. Indeed, one can immediately find that
\begin{align*}
\mathcal{W}_{\rm NJL}[J] &= N' \, \int \mathcal{D} A e^{- i \int d^4x \left[ \frac{A^2}{2G} - V(A-J) \right] } \\
&= N' \, \int \mathcal{D} A e^{- i \int d^4x \left[ \frac{(A+J)^2 }{2G} - V(A) \right] } \,, \nonumber
\end{align*}
with~$A^2= A_\mu A^\mu$ and where the one loop effective potential~$V(A)$ is obtained from 
\begin{align*}
& i \int  d^4 x V(A) = \ln {\rm det} (i \Pdd + \slash \!\!\!\! A(x) -m) - \ln {\rm det} (i \Pdd  -m) \\
&= {\rm Tr}\, \ln \frac{1}{i \Pdd-m} (i \Pdd + \slash \!\!\!\! A(x) -m) \\
&= \int d^4x\: \bigg[  {\rm Tr}\, \slash \!\!\!\! A(x) S_{F}(x,x) -\frac{1}{2} {\rm Tr} \,\int d^4 y\,\, \slash \!\!\!\! A(x) \, S_{F}(x,y)  \, \slash \!\!\!\! A(y) S_{F}(y,x) + \dots \bigg] \,,
\end{align*}
where the trace Tr denotes summation over the spinorial indices. Expanding in power series of~$A$, keeping only quadratic and quartic terms, entail the contributions from the lowest orders or the vacuum polarization integral.  
\begin{align*}
\int d^4 x V^{(2)}(x)
&= \frac{1}{2} \int \frac{d^4 q}{(2\pi)^4} \, \tilde{A}_\mu (q)  \, \tilde{A}_\nu (q) \, \left( q_\mu q_\nu  - g_{\mu \nu} q^2 \right) \Pi(q^2) \\
&= \frac{1}{2} \int \frac{d^4 q}{(2\pi)^4} \, \tilde{A}_\mu (q)  \, \tilde{A}_\nu (q) \, \left( q_\mu q_\nu  - g_{\mu \nu} q^2 \right) \frac{1}{e^2(q)} \\
&\simeq \frac{1}{4 e^2(0)} \int d^4 x F_{\mu \nu} (x) F^{\mu \nu} (x)\,,
 \end{align*}
where~$\Pi(q^2)$ is recovered from the vacuum polarization integral in quantum electrodynamics 
\begin{eqnarray}
\Pi(q^2)=\frac{1}{12\pi^2} \ln \frac{\Lambda^2}{q^2} 
\,,
 \end{eqnarray}
where a momentum cut-off~$\Lambda$ has been introduced. This entails the running of the coupling constant 
\[ \frac{1}{e(q^2)}=\frac{1}{12\pi^2} \ln \frac{\Lambda^2}{q^2} \:. \]

The effective action for the vector condensate due to quadratic term is finally provided by the expression
\begin{equation} \label{Affect}
\mathcal{S}^{(2)}= \int d^4 x\, \mathcal{L}^{(2)}(x) = \int d^4x \left[ \frac{1}{4 e^2} F_{\mu \nu} F^{\mu \nu} + \frac{A^2}{2 G} \right]\,.
\end{equation}
It is relevant also to notice that, within this framework, condensation is a by-product of a spontaneous breakdown of the Lorentz symmetry, as the condensation value of~$A_\mu$ survives when the current~$J_\mu$ is set to zero. Quartic and higher terms for~$V(A)$ will be also generated and shall be added to~\eqref{Affect}, potentially entailing a potential in~$A$ for a spontaneous symmetry breaking. 

We shall at this point comment on the fact that the spontaneous Lorentz symmetry breaking may leave potential non-covariant effects, as well as violations of the gauge invariance. Not to be the ill-defined, the theory must be therefore restricted to low energies. On the other hands, the low-energy limit will represent an equilibrium limit in a
renormalization group
flow perspective, with higher order terms becoming relevant at higher energies. The energy reference scale for these considerations would be dictated by the self-interaction coupling constant of fermions, $G$. The running of this coupling constant induces the definition of theories at the boundaries. The theory at equilibrium, in the infrared limit, can be then formally extended up to high energies, but at the cost of considering the loop corrections.

The relation between the regularization scheme in causal fermion systems and in this condensation mechanism may solve sole of the problems of consistency of this mechanism proposed by Bjorken. At the same time, one may extend the argument so to account for both Yang-Mills fields and gravity --- with the caveat of evading the Weinberg-Witten theorem.

\item
We finally comment on another perspective, accounting for unification of gravity with other forces. Ultimately, this perspective could be combined with the one referring to condensation that we spelled out in the previous point.

For this purpose, we take into account the gravi-weak unification paradigm proposed in~\cite{Alexander_2014}. This is an extended BF theory with~$\text{GL}(2, \mathbb{C})_L\times \text{GL}(2, \mathbb{C})_R$ symmetry group, which in spinorial indices~$A,B=0,1$ reads
\begin{align*}
&\mathcal{S}_{\rm GW} =\int \frac{i}{4\pi G} \Big[ B^{AB} \wedge F_{AB} - B^{A'B'} \wedge F_{A'B'}  + \frac{\lambda}{6G}   \left( B_{AB} \wedge B^{AB} - B_{A'B'} \wedge B^{A'B'}   \right) \nonumber \\
&-\frac{1}{2} \Psi_{ABCD} \, B^{(AB} \wedge B^{CD)} + \frac{1}{2} \Psi_{A'B'C'D'} \, B^{(A'B'} \wedge B^{C'D')}- \Psi_{A'B'AB} \, B^{A'B'} \wedge B^{AB} \Big] \nonumber \\
&+ \frac{i g^2}{2} \left( \Psi^2_{ABCD} + \Psi^2_{A'B'C'D'} + \Psi^2_{ABC'D'} \right)  \left( B_{AB} \wedge B^{AB} - B_{A'B'} \wedge B^{A'B'}   \right)\,,
\end{align*}
where primed and unprimed indices labeling variables in the two symmetry sectors, namely~$\text{GL}(2, \mathbb{C})_L$ and~$\text{GL}(2, \mathbb{C})_R$, are obtained from the Infeld--van der Waerden map~$e^a \rightarrow e^{AA'}=e^a \, \sigma^{AA'}_a$, involving the the Infeld--van der Waerden symbols~$\sigma^{AA'}_a$; each~$\text{GL}(2, \mathbb{C})$ connection~$A^{ab}$, with~$a,b=1,\dots 4$, rewrites according to
\[ A^{ab}= A^{AA'BB'}= \epsilon_{AB} A^{A'B'} +  \epsilon_{A'B'} A^{AB} \:, \]
and consequently for the field strength of each~$\text{GL}(2, \mathbb{C})$ connection~$A^{ab}$ it holds
\[ F^{ab}= F^{AA'BB'}= \epsilon_{AB} F^{A'B'} +  \epsilon_{A'B'} F^{AB} \:, \]
where~$F_{AB}=dA_{AB}+A_A^{\ C} \wedge A_{CB}$ denotes the field strength of~$A^{AB}$, and similar expression follows for~$F_{A'B'}=dA_{A'B'}+A_{A'}^{\ C'} \wedge A_{C'B'}$ in terms of~$A^{A'B'}$; $B^{AB}$ and~$B^{A'B'}$ are, in four dimensions, two-forms that are valued in the algebra of~$\text{GL}(2, \mathbb{C})_\mathbb{C}$, with indices 
\[ B^{ab}= B^{AA'BB'}= \epsilon_{AB} B^{A'B'} +  \epsilon_{A'B'} B^{AB} \:; \]
the multiplets of scalar fields have been introduced, as Lagrangian multipliers, which in the Lorentzian indices read~$\Psi_{abcd}$, and in the spinorial indices decompose into pure spin-2 fields, i.e.\ $\Psi_{ABCD}$ and~$\Psi_{A'B'C'D'}$, totally symmetric, and~$\Psi_{ABA'B'}$, with mixed components and symmetric pairs of indices; the dimensionless cosmological constant~$\lambda=G\Lambda$ and dimensionless coupling constant~$g$ have been introduced.

An explicit solution that breaks the left-right symmetry has been recovered in~\cite{Alexander_2014}, which simultaneously provides gravity on one side, and the electroweak interactions on the other side provided with parity violation. Indeed, at the lowest order of the symmetry breaking solution, one obtains 
\begin{eqnarray} \label{grave}
\mathcal{S}^{(0)}=\!\!\!\!\!\!\!\!\!&&\int \frac{i}{4 \pi G} \Sigma^{AB} \wedge F_{AB} + \frac{\lambda}{12 \pi G^2} e - \frac{e}{4 g^2_{\rm YM}} F^{A'B'}_{\mu \nu} F_{A'B' \rho \sigma} g^{\mu \rho} g^{\nu \sigma} \nonumber\\
&&- i \Theta F^{A'B'} \wedge F^{A'B'} + \frac{9G^2}{(16\pi)^2 \lambda^2 e}(F_{(A'B'}\wedge F_{C'D')})^2\,,
\end{eqnarray}
with~$\Sigma^{AB}=e^A_{\ C'} \wedge e^{B C'}$, $F_{AB}=R_{AB}$ field strength of the spin-connection, and~$\Theta$ and~$g^2_{\rm YM}$ known functions of both~$\lambda$ and~$g^2$. The Yang-Mills sector is parity violating, as suggested by the appearance of the Pontryagin term.

The inspection of the first order correction in~$g^2$ to~\eqref{grave} provides the existence of terms such as
\begin{align*} \label{grawi}
\mathcal{S}^{(1)}=
\int \frac{i g^2}{4 \pi G} \bigg[& b^{AB} \wedge F_{AB} + \frac{\lambda}{3G} b_{AB} \wedge \Sigma^{AB} \nonumber\\
&+  \frac{\lambda g^2}{3G} b_{AB} \wedge b^{AB} +  \frac{\lambda g^2}{3G} b_{A'B'} \wedge b^{A'B'} \bigg] 
+ \dots \,, 
\end{align*}
with~$b^{AB}$ and~$b^{A'B'}$ auxiliary fields determined by variation of~$\mathcal{S}^{(1)}$. The~$b^{AB}$, in particular, are responsible for the emergence of parity violating effects also in the gravitational sector: varying with respect to~$b^{AB}$ provides the gravitational Pontryagin density, which is not multiplied by other evolving fields.
\end{enumerate}

This proposal for gravi-weak unification resonates strongly with the $E_8 \times E_8$ unification proposed in~\cite{Kaushik}. In the latter work, spontaneous chiral symmetry breaking gives rise to the electroweak sector $SU(2)_L\times U(1)_Y \rightarrow U(1)_{em}$, and also to a `darkelectro-grav' sector $SU(2)_R \times U(1)_{YDEM}\rightarrow U(1)_{DEM}$. The broken $SU(2)_R$ symmetry has been proposed as the origin of general relativity, and the unbroken DEM (Dark Electromagnetism) symmetry has been suggested~\cite{mond} to be the origin of Milgrom's Modified Newtonian Dynamics (MOND). 

\appendix
\section{Description of Majorana Spinors in Causal Fermion Systems} \label{appmajorana}
In some applications of octonions to the standard model, it is essential or helpful to
describe the neutrinos by Majorana spinors (see for example~\cite{singh-octo1, Addazi:2019dqt}).
In order to facilitate the comparison of the different approaches, we here
explain how Majorana spinors can be described in the setting of causal fermion systems.

In the setting of causal fermion systems, one cannot work with two-component Weyl spinors.
The reason is that, in order to obtain the spinor bundle from a causal fermion system,
it is crucial that the isometry group of the spin spaces (with respect to the spin inner product)
contains the usual spin group (for details see for example~\cite{topology}).
This condition is fulfilled for four-component Dirac spinors, where the isometry group of
the spin spaces is~$\U(2,2) \supset \Spin(1,3)$ (the physical significance of the
corresponding local $\U(2,2)$-transformations is explained in~\cite{u22}).
However, for two-component Weyl spinors, the group~$\text{SL}(2,\C)$ describing the Lorentz transformations
cannot be realized as the isometry group of a two-dimensional spin space
(no matter how the signature of the inner product on the spin space is).

With this in mind, in the setting of causal fermion systems it is most natural to work with 
four-component Dirac spinors. But it is also possible to describe two-component spinorial equations,
as we now describe. In the {\em{massless}} case, the Dirac equation decouples into two two-component
equations describing the left- and right-handed components.
Setting for example the right-handed component to zero, one gets a massless equation for Weyl spinors,
which can be written as
\[ \chi_R \,\big( i \gamma^j \partial_j \big)\, \chi_L \Psi = 0 \:. \]
This method is used for the description of left-handed massless neutrinos in~\cite{pfp, cfs}.
In the massive case, however, the left- and right-handed components are coupled.
This is why in~\cite{cfs} massive neutrinos have a left- and right-handed component,
which together form a four-component Dirac spinor (just as in the sectors describing the
quarks and the charged leptons). The right-handed components
couple only to the gravitational field and are sometimes referred to as {\em{sterile neutrinos}}.

Another method for reducing the number of spinorial degrees of freedom is to impose that
the wave functions satisfy a local symmetry, which must be compatible with the Dirac equation.
Implementing this method by imposing that in a specific spinorial basis the
wave functions be real-valued gives rise to {\em{Majorana spinors}}, as we now explain.
We work with the Dirac matrices in the Majorana representation
\begin{align*}
\gamma^0 &= \begin{pmatrix} 0 & -\sigma^2 \\ -\sigma^2 & 0 \end{pmatrix} \:,&
\gamma^1 &= \begin{pmatrix} 0 & i \sigma^3 \\ i \sigma^3 & 0 \end{pmatrix} \:,\\
\gamma^2 &= \begin{pmatrix} i\,\1_{\C^2} & 0 \\ 0 & -i\,\1_{\C^2} \end{pmatrix} \:,&
\gamma^3 &= \begin{pmatrix} 0 & -i \sigma^1 \\ -i \sigma^1 & 0 \end{pmatrix} \:.
\end{align*}
where~$\sigma^\alpha$ with~$\alpha=1,2,3$ are the three Pauli matrices.
The spin inner product (with respect to which the Dirac matrices are symmetric) takes the form
\[ \Sl .|. \Sr = \la \,.\,, \gamma^0 \,.\, \ra_{\C^4} \:. \]
The pseudo-scalar matrix becomes
\[ \gamma^5 = \begin{pmatrix} 0 & i\,\1_{\C^2} \\ -i\,\1_{\C^2} & 0 \end{pmatrix} \:. \]
We now consider the Dirac operator with a mass~$m$ and pseudo-scalar mass~$n$,
\beq \label{dirgen}
\big( i \gamma^j \partial_j + i \gamma^5 n - m \big) \psi = 0 \:.
\eeq
Choosing both masses to be real, one verifies directly that all the matrix entries on the left are real.
Therefore, the equation admits real-valued solutions, i.e.\
\beq \label{majorana}
\psi(x) \in \R^4 \:.
\eeq
Restricting attention to solutions of this form, the Dirac equation reduces to the Majorana equation.
Clearly, the Majorana equation can be written equivalently in the usual way as a two-component equation.
But this is only a matter of convenience. In the context of causal fermion system, it is preferable
to describe Majorana spinors by the four-component equation~\eqref{dirgen} under the constraints~\eqref{majorana},
because in this way, we can retain the four-dimensional spin space endowed with an inner product
of signature~$(2,2)$. The Lorentz transformations of the Majorana spinors are described by
the subgroup of~$\U(2,2)$ generated by the bilinear matrices (which, in the Majorana representation,
again have real-valued entries).

We finally note that the reality condition~\eqref{majorana} greatly restricts the space of solutions.
In particular, it is impossible to build up Dirac sea configurations (in simple terms because~\eqref{majorana}
rules out plane-wave solutions~$\sim e^{i \omega t}$). But there are interesting
Lorentz invariant solutions, like the two-point distribution
\[ p_m(x,y) := \int \frac{d^4k}{(2 \pi)^4}\: (\slashed{k} + \gamma^5 n + m)\: \delta\big( k^2 - n^2-m^2 \big)\:
e^{-i k (y-x)} \:. \]
It is an open problem whether regularizing such distributions gives rise to minimizers of the causal action principle
which could replace the Dirac sea configurations in the analysis of the causal action principle.

\Thanks{{{\em{Acknowledgments:}} 
We are grateful to the ``Universit\"atsstiftung Hans Vielberth'' for generous support.}

\providecommand{\bysame}{\leavevmode\hbox to3em{\hrulefill}\thinspace}
\providecommand{\MR}{\relax\ifhmode\unskip\space\fi MR }
\providecommand{\MRhref}[2]{%
  \href{http://www.ams.org/mathscinet-getitem?mr=#1}{#2}
}
\providecommand{\href}[2]{#2}

\end{document}